\documentclass[review,12pt,authoryear]{elsarticle}

\usepackage{amsmath,amssymb,amsfonts}
\usepackage{graphicx}
\usepackage{textcomp}
\usepackage{array}
\usepackage{multirow}
\usepackage{float}
\usepackage{enumitem}
\usepackage{subcaption}
\usepackage{algorithm}
\usepackage{algpseudocode}
\usepackage{hyperref}
\usepackage{cleveref}
\usepackage{caption}

\journal{Transportation Research Part C}
\begin{document}

\begin{frontmatter}

\title{Trajectory-Based Optimization for Air Traffic Control in the Terminal Maneuvering Area}

\author[inst1]{Yutian Pang\corref{mycorrespondingauthor}}
\cortext[mycorrespondingauthor]{Corresponding author.}
\ead{yutian.pang@austin.utexas.edu}
\author[inst2]{Daniel Delahaye}
\author[inst1]{John-Paul Clarke}

\affiliation[inst1]{organization={Department of Aerospace Engineering and Engineering Mechanics, The University of Texas at Austin},
            city={Austin},
            postcode={78712},
            state={TX},
            country={USA}}

\affiliation[inst2]{organization={Optimization and Machine Learning Team, Ecole Nationale de l'Aviation Civile (ENAC)},
            city={Toulouse},
            postcode={31055},
            country={France}}

\begin{highlights}
\item We derive a closed-form trajectory-based formulation of the three-segment aircraft terminal arrival path as a smooth nonlinear function of the base-leg extension, which directly couples path geometry to arrival time and enables the optimization that realize the required landing separation through terminal control actions.
\item We implement and evaluate three landing-sequence policies, First-Entry-First-Serve (FEFS), First-on-Final-First-Serve (FOFFS), and FOFFS with Constrained Position Shifting (CPS) up to $k$ positions, realized by mixed-integer linear programming (MILP) and nonlinear programming (NLP).
\item We considers a heterogeneous mixed weight class fleet with pair-specific wake-turbulence separation threshold, type-specific runway occupation, per-aircraft final-approach speed bounds, as well as wind uncertainty modeled by a Gaussian distribution and and projected onto the aircraft heading to convert commanded airspeeds into ground speeds.
\item We conduct large scale Monte Carlo evaluation spanning various arrival densities across all scheduling policies and a  discretization grid, with real-time solver-runtime analysis demonstrating near real-time per-entry performance.
\end{highlights}

\begin{abstract}
We present a trajectory-based optimization framework for arrival sequencing and scheduling in the terminal maneuvering area (TMA). In contrast to node-link scheduling models that abstract trajectories into time-delay parameters, the proposed method computes implementable per-aircraft speed profiles and path extensions that realize the required landing separation through terminal air traffic control actions. The framework couples an analytic TMA path model, comprising a tangent leg, a radius-to-fix turn, and a final-approach segment, with a nonlinear program (NLP) that jointly optimizes path stretch and segment speeds under a weighted objective. Three landing-order policies are evaluated. First-Entry-First-Serve (FEFS), First-on-Final-First-Serve (FOFFS), and FOFFS with Constrained Position Shifting (CPS) up to $k$ positions, realized by a two-phase scheme coupling mixed-integer linear programming (MILP) with NLP that selects an optimized landing order before trajectory optimization. The aircraft population is drawn from a realistic weight class fleet mix with pair-specific wake-turbulence separation, and each scenario is perturbed by a Gaussian wind sample, projected per segment onto the aircraft heading to convert commanded airspeeds into ground speeds. Moreover, an online, rolling-horizon formulation commits each aircraft's trajectory irrevocably upon entry, enabling real-time decision-making. Monte Carlo experiments on the simplified A80 TMA show that: (i) FOFFS consistently outperforms FEFS in delay, path stretch, and fuel burn by exploiting geometric asymmetries among arrival streams; (ii) CPS further reduces separation violations and path stretch, with diminishing returns and a fast-growing solver cost; (iii) fuel estimates from BADA~3 and OpenAP agree on the qualitative trends; and (iv) the per-entry optimization completes in near real-time, demonstrating real-time feasibility for practical deployment.
\end{abstract}


\end{frontmatter}


\section{Introduction}
\label{sec:intro}
The sustained growth of global air traffic demand continues to intensify congestion and delay across the National Airspace System (NAS), and the terminal maneuvering area (TMA) has increasingly become the dominant bottleneck that limits end-to-end throughput. In metroplex environments such as in Atlanta and Denver, multiple arrival and departure streams interact within tightly constrained terminal airspace, requiring precise synchronization to maintain separation standards while sustaining runway utilization \citep{ren2009contrast, balakrishnan2006scheduling, clarke2013optimized}. Under rising demand, First-Come-First-Served (FCFS) operations become inefficient, because delay absorption is frequently achieved through tactical vectoring, path stretching, or holding, all of which can increase fuel burn, noise exposure, and unrecoverable delay \citep{ng2024optimization}. These challenges motivate the development of decision-support tools that shift terminal operations from reactive, clearance-by-clearance control toward trajectory-based tactical planning and execution in a dynamic fashion.

Air traffic control (ATC) services in the United States are organized into three layers: (i) en-route control, provided by Air Route Traffic Control Centers (ARTCCs); (ii) terminal radar approach control (TRACON), which manages arrivals and departures within the TMA (i.e., typically a 30--40 nautical-mile radius airspace volume below 10,000 feet); (iii) airport tower control, responsible for the immediate runway vicinity \citep{de2020airport,bennell2011airport}. For arriving traffic, the TRACON controller's tactical objective is to transform stochastic inbound streams (entering from multiple feeder gates) into an orderly, safely separated flow delivered to the Final Approach Fix (FAF) and runway threshold, at which point the aircraft is handed off to the final approach and tower controllers \citep{balakrishnan2006scheduling}. This objective again couples three interacting elements: (i) runway acceptance capacity, governed by separation minima including wake-turbulence constraints; (ii) the evolving geometry of converging arrival flows; and (iii) the controller's available tactical actions such as the usage of speed adjustments, path stretching or shortening (e.g., downwind/base-leg extensions, holding loops, shortcuts). The resulting arrival sequencing and scheduling task has a long history in the operations research literature and remains computationally and operationally challenging, because solutions must be generated quickly, remain robust to uncertainty, and be acceptable to controllers and stakeholders \citep{bennell2011airport}.

A major strand of terminal-area automation in the U.S.\ emerged from the Center-TRACON Automation System (CTAS), which integrated trajectory prediction tools such as the Traffic Management Advisor, the Descent Advisor (DA), and the Final Approach Spacing Tool (FAST) to provide controller advisories for runway assignments, landing sequences, and terminal-area spacing \citep{denery1995center, halverson1992systems}. FAST, in particular, was designed to generate speed and heading advisories that map directly to terminal control actions, supported by aircraft performance models and wind modeling in the trajectory prediction engine. In parallel, a noise-abatement research thread developed the methodology and simulation tooling needed to design and analyze Continuous Descent Arrival (CDA) procedures \citep{clarke1997systems, ren2007modeling, lowther2008enroute, clarke2013optimized, cao2011evaluation}. Clarke's NOISIM framework \citep{clarke1997systems} established the coupling of trajectory simulation and noise impact; \citet{ren2007modeling} extended this line with a Monte Carlo fast-time simulator and a separation-analysis methodology that accounts for inter-flight wind variations and FMS behavior, and validated the approach on the KSDF CDA flight test; \citet{lowther2008enroute} further investigated en-route speed optimization for CDA; and \citet{cao2011evaluation} evaluated CDA as a standard terminal operation and explicitly integrated a CDA scheduling step with conflict resolution. These systems improved situational awareness and reduced delays through time-based metering, advisory guidance, and trajectory-aware sequencing. However, practical terminal operations may still rely on static spacing intervals and manual tower--TRACON coordination, which can compromise throughput when arrivals and departures are interdependent~\citep{diffenderfer2013automated}. More broadly, many advisory and scheduling systems provide target times or sequences without explicitly generating the detailed, geometry-feasible speed and path controls required to realize those targets under realistic TRACON constraints.

In the literature, a widely cited synthesis of runway scheduling research classifies problems into landing scheduling, departure scheduling, and runway sequencing, and catalogs solution approaches ranging from exact methods (e.g., dynamic programming, mixed-integer programming, branch-and-bound) to heuristics and meta-heuristics \citep{bennell2011airport}. A foundational theme in practical arrival scheduling is the tradeoff between throughput efficiency and fairness. The FCFS discipline is operationally simple but can be suboptimal, motivating constrained position shifting (CPS), first proposed in \cite{dear1976dynamic}, in which each aircraft's optimized position may deviate by at most a small number of places from its FCFS order to balance equity with performance \citep{balakrishnan2006scheduling}. \citet{psaraftis1980dynamic} gave a dynamic-programming algorithm for this problem, and \citet{balakrishnan2006scheduling} later developed a network-based DP that scales linearly in the number of aircraft and handles pair-specific wake separation minima, arrival-time windows, and same-route precedence constraints explicitly. MILP formulations of the same problem in \cite{beasley2000scheduling} established the standard big-M disjunctive representation used throughout the subsequent literature. Another line of work extends the optimization boundary beyond the runway to the TMA itself, framing the Terminal Airspace Scheduling Problem (TASP) as a joint sequencing and scheduling problem over a node-link network of waypoints and route segments \citep{capozzi2009optimal, desai2016optimization, ng2024optimization}. More recent formulations explicitly incorporate TMA decisions (e.g., speed control, vectoring, holding, and structured merges) alongside runway sequencing, reporting meaningful delay reductions at busy airports relative to FCFS \citep{gui2025metaheuristic, dhief2023meta}. Other related problems include runway operations planning for departures and crossings \citep{anagnostakis2003runway} and integrated arrival-departure coordination with runway dependent constraints \citep{liang2017integrated}.

In parallel with runway and TMA scheduling research, both NextGen and SESAR have adopted Trajectory-Based Operations (TBO) as the operating paradigm of their future airspace system \citep{torres2012integrated, gardi2018multiobjective}. The core idea of TBO is that each flight is managed as a four-dimensional trajectory contract, negotiated between airborne flight-management systems and ground decision-support systems, rather than as a loose sequence of clearances. Expected benefits include more predictable traffic flows, fuel- and time-efficient descents, reduced controller workload, and earlier tactical conflict resolution \citep{torres2012integrated, gardi2018multiobjective, clarke2013optimized}. However, realizing the runway-level benefits of TBO requires scheduling logic that can reason over, and produce, physically feasible trajectories rather than merely time-over-waypoint targets. Otherwise, the trajectories produced by the airborne FMS and the schedules produced by ground-based tools are not guaranteed to be mutually consistent, and the resulting deconfliction loops erode the anticipated benefits \citep{gardi2018multiobjective}.

Many authors formulate the terminal airspace scheduling problem (TASP) as a node-link flow problem \citep{capozzi2009optimal, desai2016optimization, ng2024optimization, dhief2023meta, gui2025metaheuristic}. In these formulations, each arrival route is modeled as a chain of nodes (waypoints, holding fixes, merge points) connected by edges whose travel-time envelopes are parameterized by minimum/maximum speeds or a set of precomputed profiles. The optimizer then selects time-over-node variables subject to edge travel-time bounds and separation at the runway. This abstraction has been shown to be sufficient for large TMA networks and for integrating operational interventions such as holding stacks, vectoring and point-merges \citep{ng2024optimization, dhief2023meta}, but it has two structural limitations. First, it does not compute aircraft trajectories explicitly (the TASP does not model aircraft trajectories explicitly \citep{ng2024optimization}), which means the schedule is agnostic to whether a given time-over-node assignment is actually flyable by a given aircraft type from a given entry heading. Second, the node-link abstraction hides the geometric coupling between path stretch and arrival time. In real terminal operations, base-leg extension is a continuous control action whose effect depends on the aircraft's current position and the turn radius, and it cannot in general be represented as a single edge with a linear or piecewise-linear travel-time function. Our work complements this direction by making the opposite abstraction choice. We give up the flow-network generality in exchange for an analytic closed-form trajectory geometry that directly exposes $d_i$ (base-leg extension) and the segment-wise speeds as decision variables, and that produces implementable speed profiles and FAF-extension waypoints as its output. Against this background, we focus on the central motivation of this paper, \textit{how can an automated terminal sequencing and scheduling capability be formulated so that its output is not only a runway-time schedule, but also a set of feasible, controller-usable aircraft trajectories that realize the schedule through realistic terminal control actions?}

This paper addresses the above question by formulating a trajectory-based optimization framework for terminal arrival sequencing and scheduling in TRACON operations, with a current focus on single-runway arrivals. Instead of node-link abstractions, the proposed method computes implementable arrival trajectories that achieve the required landing spacing through a set of tactical controls: path stretching and speed control. Vectoring-for-spacing via repeated S-turns and holding patterns is not considered at the moment, due to an increased airspace complexity and complicated controller and pilot intent management \citep{artiouchine2008runway, bennell2011airport}. Nominal estimated arrival times are computed from simplified aircraft kinematics, which directly couples scheduling decisions to feasible trajectory construction. Following standard FCFS conventions on shared jet routes \citep{neuman1991analysis, balakrishnan2006scheduling}, the landing order of aircraft on the same arrival route is preserved. To evaluate robustness across demand conditions, the arrival-generation process is modeled as a shifted Poisson process, consistent with the near-exponential inter-arrival distributions reported at major U.S. airports \citep{willemain2004statistical}. We highlight the following contributions here,
\begin{itemize}
    \item Our primary technical contribution is an analytic, closed-form expression of the three-segment terminal arrival path (tangent leg, radius-to-fix arc, and final straight-in segment) as a function of the base-leg extension $d_i$, the aircraft entry coordinates, and the FAF location. The derivation uses a vector-product formulation where the tangent point on the RF turn circle is written in closed form using the center-to-entry vector $\mathbf{v}_i$ and its perpendicular $\mathbf{v}_i^\perp$, the RF arc angle $\theta_i(d_i)$ is obtained from the cross- and dot-products of the two radius vectors at the turn center, and the total flight time becomes a smooth nonlinear function of $d_i$ alone.
    \item Prior node-link TMA formulations typically assume identical aircraft or separation into a single scalar, we formulate terminal arrival scheduling as a trajectory-based operation in which the output is a segment-wise commanded speed profile together with final-leg extensions that can serve as auxiliary waypoints in the terminal arrival procedure. We model a mixed Heavy/Large/Small fleet with aircraft-specific reference landing speed $V_{\mathrm{ref}}$ and a pairwise wake-turbulence separation matrix, which exposes the relative benefit of constrained resequencing.
    \item Rather than scheduling arrivals in static batches, we determine each arriving aircraft's trajectory in near real-time from the evolving arrival set. The dynamic formulation uses the current entering aircraft together with a lookahead window of preview aircraft, and each aircraft is \emph{touched only once}, meaning the trajectory is committed irrevocably upon entry into the TRACON. On top of the FOFFS baseline, we formulate a FOFFS-CPS policy in which a lightweight mixed-integer linear program (MILP) \citep{dear1976dynamic, balakrishnan2006scheduling} but embedded in a rolling online window that selects an optimized landing order that is allowed to deviate by at most $k$ positions from the FOFFS rank, followed by the same trajectory NLP. The decomposition retains the tractability of entry-order scheduling while recovering most of the benefit of a fully optimal resequencing, and it allows us to view the safety--throughput trade-off when $k$ varies.
    \item Practical terminal operations are subject to winds that cause ATC-commanded airspeeds to differ from the ground speeds that actually drive arrival times. We incorporate a uncertainty model in which a scalar wind is sampled from $\mathcal{N}(\mu_w,\sigma_w^2)$ for each scenario and projected per segment onto the heading direction using the straight-leg headings and the chord direction of the RF turn~\citep{ren2007modeling}. The resulting segment-wise ground speeds drive both the NLP arrival-time constraint and the MILP window bounds, allowing the Monte Carlo study to characterize operational uncertainty without reformulating the optimization as a chance-constrained program.
    \item We evaluate five policies FEFS, FOFFS, FOFFS-CPS$_{k\in\{1,2,3\}}$, under a $2\times 2$ grid of discretization parameters (e.g., $\Delta_d\in\{0.5,1.0\}$ nmi, $\Delta_s\in\{5,10\}$ kts to reflect real world air traffic control options) with $1,000$ demand seeds and $10$ wind samples per seed, yielding nearly $200,000$ per-run records. Beyond the usual safety, delay, path-stretch, and fuel-burn metrics, we also report the per-entry solver runtime, which determines whether the scheme can run at the rate at which aircraft enter the TRACON. This analysis is essential for positioning the framework as a real-time decision-support tool rather than an offline scheduler.
\end{itemize}

\begin{figure}[t]
    \centering
    \includegraphics[width=\textwidth]
    {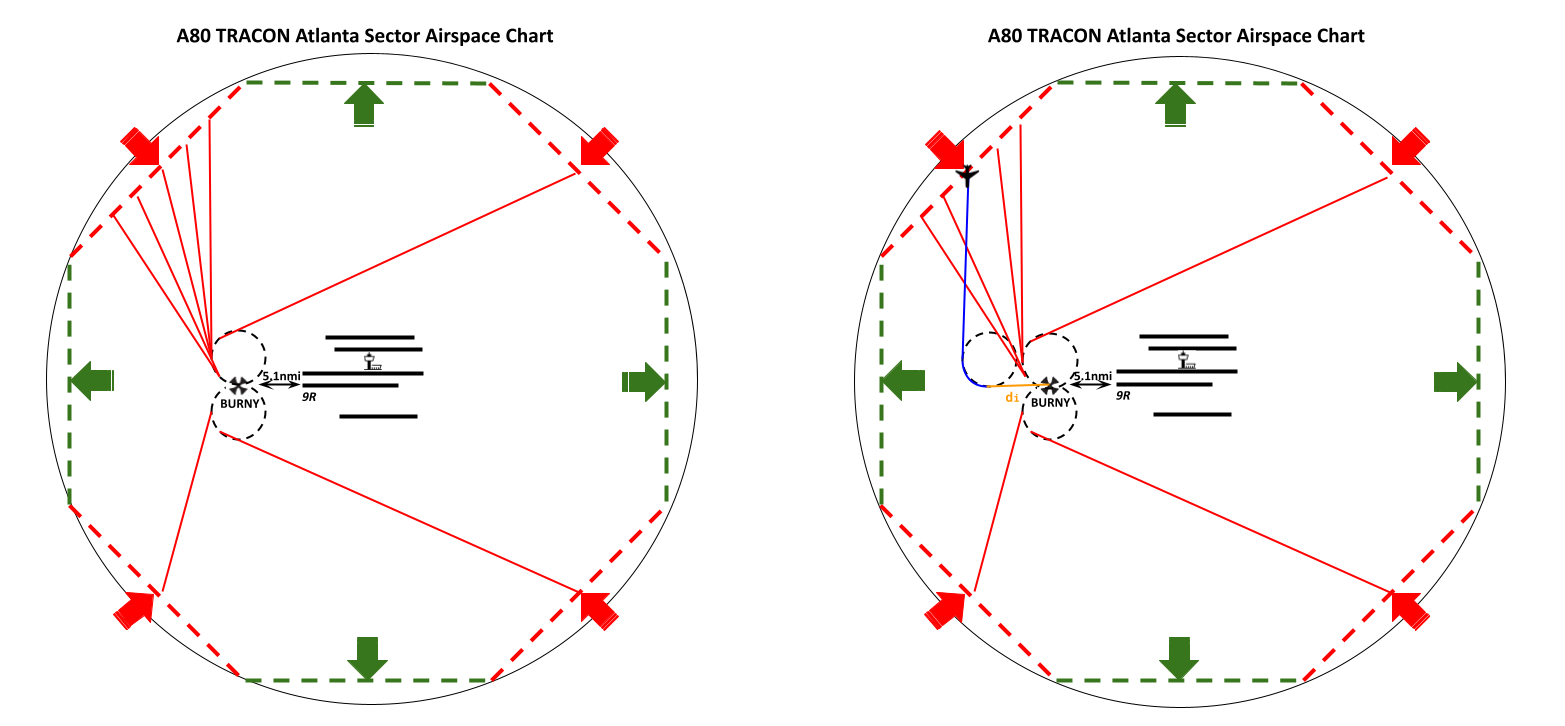}
    \caption{Schematic diagram of the Atlanta VORTAC 30 DME (TCP) of the Atlanta Large TRACON (A80), where the A80 traffic controller takes over arrivals from the Atlanta Air Route Traffic Control Center (ZTL). The circular airspace approximately represents the A80 ATL Sector approach-control boundary, with the four feeder-gate arrival procedures highlighted in red. In this study, all arrivals are targeting a westerly landing on runway 9R at KATL, with BURNY as the Final Approach Fix (FAF). The left panel shows the ideal case in which each arriving aircraft follows its shortest path and a single radius-to-fix (RF) turn to the FAF. In practice (right panel), each aircraft may instead be routed through an extended FAF auxiliary waypoint at a distance $d_i$ upstream of BURNY to ensure adequate separation at the FAF.}
    \label{fig: A80}
\end{figure}

\section{Methodology \label{sec: methodology}}
In this section, we present the proposed trajectory-based optimization framework. As illustrated in \Cref{fig: A80}, each aircraft extends its base leg by a variable distance $d_i$ upstream of the FAF. This extension provides precise temporal spacing without fuel-inefficient holding patterns or vector-for-spacing.

The framework has five main components. First, a mixed-fleet traffic generation process with pair-specific wake-turbulence separation is defined over four corner feeder gates in \Cref{subsec: flow_generation}. Second, the trajectory geometry (\Cref{subsec: travel_time}) computes arrival times from path extensions and segment speeds via the analytic TRACON path model, and the closed-form geometry is extended to account for a per-segment wind correction derived from wind sample in \Cref{subsec: wind}. Third, we define three landing-sequence policies, FEFS, FOFFS, and FOFFS with CPS, the last of which is obtained through a two-phase online sequencer coupling a MILP ordering step with the trajectory NLP in \Cref{subsec:landing_order}. Fourth, the trajectory optimization itself is formulated as a nonlinear program (NLP) that jointly determines path extensions and per-segment speeds to satisfy pair-specific separation while minimizing a weighted safety, throughput, delay, efficiency objective (\Cref{subsec:baseline_nlp}), and embedded into an online, rolling-horizon loop that commits each aircraft upon TRACON entry in \Cref{subsec:online}. Fifth, a BADA-consistent fuel consumption model (\Cref{subsec:fuel_consumption}) is used together with the OpenAP model to cross-validate fuel estimates.

\subsection{Generation of Arrival Flows \label{subsec: flow_generation}}
We simulate arrival traffic originating from four distinct corner fixes surrounding the terminal airspace: DALAS (northwest), LOGEN (northeast), HUSKY (southeast), and TIROE (southwest). Each corner fix generates an independent stream of aircraft modeled by a shifted Poisson process, as in our previous work \citep{pang2025modeling}. This approach incorporates the stochastic nature of arrival demands while strictly enforcing the minimum separation standards required for safety-critical air traffic operations.

\subsubsection{Shifted Poisson Arrival Process}
We denote the set of corner fixes as $\mathcal{K} = \{\text{DALAS},\allowbreak \text{LOGEN},\allowbreak \text{HUSKY},\allowbreak \text{TIROE}\}$. For each corner $\kappa \in \mathcal{K}$, the arrival times are generated as a renewal process where the inter-arrival time $\Delta t$ consists of a deterministic safety buffer and a stochastic component. Let $T_{\mathrm{sep}}$ denote the minimum required separation. The inter-arrival time is defined as,
\begin{equation}
\label{eq:interarrival}
\Delta t = T_{\mathrm{sep}} + X_\kappa, \quad X_\kappa \sim \text{Exp}(\lambda_\kappa)
\end{equation}
where $\lambda_\kappa$ is the arrival rate for corner $\kappa$ (in aircraft per hour). The entry time of the $m$-th aircraft at corner $\kappa$, denoted $\tau_{\kappa,m}$, is generated recursively:
\begin{equation}
\label{eq:recursive_entry}
\tau_{\kappa,m} = \tau_{\kappa,m-1} + T_{\mathrm{sep}} + \text{Exp}(\lambda_\kappa), \quad \tau_{\kappa,0} = 0
\end{equation}
This formulation ensures that no two aircraft from the same stream enter the simulation closer than $T_{\mathrm{sep}}$ seconds apart. The generation process terminates when the next computed entry time would exceed a maximum simulation window $T_{\max}$, ensuring all generated aircraft can be processed within the optimization horizon.

\subsubsection{Traffic Density Sampling}
To evaluate the optimization framework across a wide spectrum of congestion levels, we vary the arrival rate $\lambda_\kappa$ for each corner fix independently in every simulation scenario. Unlike previous approaches that sample rates from a log-uniform distribution, we employ a uniform integer sampling scheme to provide equal probability mass across the specified range of traffic densities.
For each corner $\kappa$, the arrival rate (in aircraft per hour) is sampled as:
\begin{equation}
\label{eq:lambda_sampling}
\lambda_\kappa \sim \mathsf{Uniform}(\lambda_{\min}, \lambda_{\max})
\end{equation}
where $\lambda_{\min} = 1$ and $\lambda_{\max} = 30$ aircraft per hour in our experiments. This discrete uniform sampling ensures that each integer rate value in the range has equal probability, facilitating straightforward interpretation of results and enabling direct comparison across scenarios with identical demand levels.

\subsubsection{Heterogeneous Fleet and Wake-Turbulence Separation}
\label{subsubsec: fleet}
Each generated aircraft is assigned a weight class and a specific aircraft type. The weight classes $\{\text{Heavy}, \text{Large}, \text{Small}\}$ are drawn with probabilities $(0.4, 0.4, 0.2)$, which is the fleet mix used in \cite{balakrishnan2006scheduling} in their benchmark CPS study and is representative of major U.S.\ hub airports (Heavy and Large carriers dominate). Within each class, the aircraft type is sampled uniformly. \Cref{tab:ac_params} lists the six aircraft types considered, together with their weight class, type-specific runway-occupation time $T_{\mathrm{rwy}}$, and reference landing speed $V_{\mathrm{ref}}$. The $V_{\mathrm{ref}}$ values are used to size the per-aircraft final-approach speed bounds (\Cref{subsec:baseline_nlp}); the runway-occupation times enter the effective separation. A heterogeneous fleet is essential for the present study because, as observed in \cite{balakrishnan2006scheduling}, a homogeneous fleet provides no CPS gain. The benefit of position shifting comes precisely from re-pairing aircraft to avoid the large Heavy followed by Small separation penalty.

The minimum time-based separation at the FAF between a leading aircraft of class $c_l$ and a trailing aircraft of class $c_t$ follows the pair-specific wake-turbulence matrix $T_{\mathrm{sep}}^{\mathrm{wake}}(c_l,c_t)$ in \Cref{tab:wake}. The effective required separation for the $k$-th landing slot is the maximum of the wake requirement and the trailing aircraft's runway-occupation time,
\begin{equation}
T_{\mathrm{sep}}^{\,k} \;=\; \max \!\bigl(T_{\mathrm{sep}}^{\mathrm{wake}}(c_{\Pi(k-1)},c_{\Pi(k)}),\; T_{\mathrm{rwy}}(\Pi(k))\bigr),
\end{equation}
which replaces the scalar $T_{\mathrm{sep}}$ in all subsequent separation constraints. This pair-specific treatment makes the benefit of constrained resequencing explicit. The marginal cost of a Heavy-Small pairing is far larger than a Large-Large pairing, so a CPS policy can reduce total delay by re-ordering nearby aircraft in the landing queue.

\begin{table}[t]
\centering
\caption{Aircraft types used in the Monte Carlo study. Each scenario samples a weight class with probabilities $(0.4,0.4,0.2)$ for (Heavy, Large, Small) following~\citep{balakrishnan2006scheduling}, and then samples a type uniformly within the class.}
\label{tab:ac_params}

\begin{tabular}{lccc}
\hline
Aircraft type & Weight class & $T_{\mathrm{rwy}}$ (s) & $V_{\mathrm{ref}}$ (kts) \\ \hline
A359 & Heavy & 85 & 140 \\
B773 & Heavy & 85 & 150 \\
A321 & Large & 66 & 140 \\
B737 & Large & 62 & 142 \\
A221 & Small & 72 & 130 \\
B735 & Small & 72 & 127 \\ \hline
\end{tabular}
\end{table}

\begin{table}[t]
\centering
\caption{Minimum wake-turbulence separation $T_{\mathrm{sep}}^{\mathrm{wake}}(c_l,c_t)$ (seconds) at the FAF between leading class $c_L$ (rows) and trailing class $c_T$ (columns).}
\label{tab:wake}
\begin{tabular}{cccc}
\hline
$c_L \backslash c_T$ & Small & Large & Heavy \\ \hline
Small & 94 & 64  & 60 \\
Large & 94 & 64  & 60 \\
Heavy & 150 & 118 & 82 \\ \hline
\end{tabular}
\end{table}

\subsubsection{Global Indexing and Geometry Initialization}
The independent streams from all four corners are aggregated to form the global set of aircraft $\mathcal{I} = \{1, \dots, n\}$ used in the optimization formulation. We sort the total $n$ aircraft by their entry times such that $\tau_1 \le \tau_2 \le \dots \le \tau_n$, where $\tau_i$ denotes the TCP entry time of aircraft $i$.
Crucially, each aircraft $i$ inherits the spatial properties of its generating corner $\kappa$. We define the entry state $C_i = (X_{C_i}, Y_{C_i})$ as the coordinates of the corner fix $\kappa$ where aircraft $i$ originated. This entry position $C_i$ determines the specific geometric flight path characteristics utilized in the travel time calculation in Section \ref{subsec: travel_time}.

\begin{figure}[!t]
    \centering
        \begin{subfigure}[t]{0.45\textwidth}
            \includegraphics[width=\textwidth]{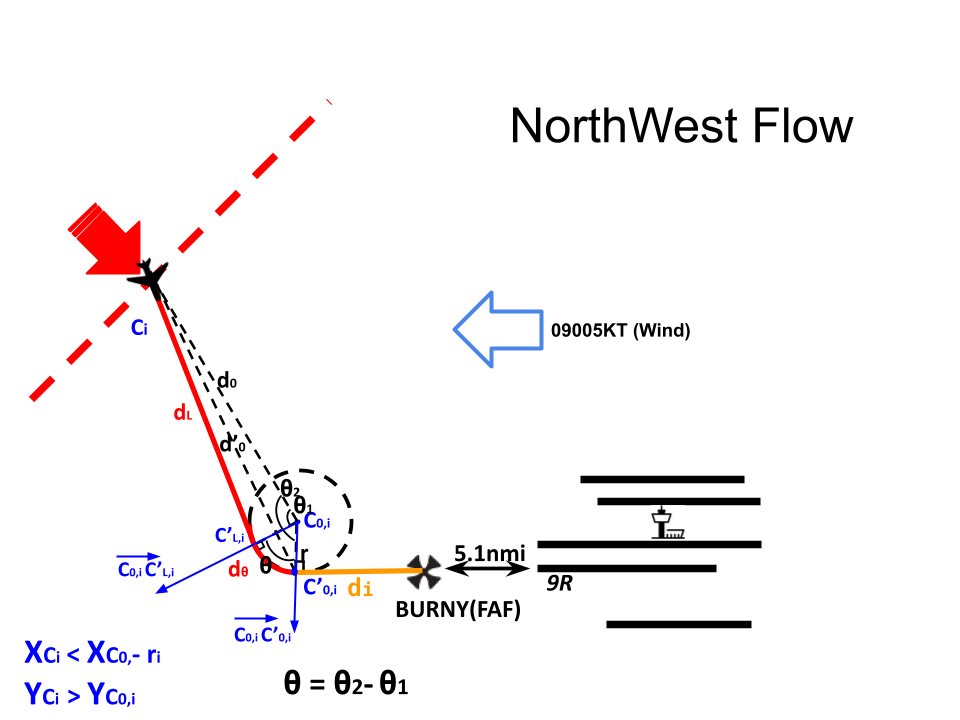}
            \caption{Landing geometry of northwest arrivals.}
            \label{fig: NorthWestFlow}
        \end{subfigure}
        \begin{subfigure}[t]{0.45\textwidth}
            \includegraphics[width=\textwidth]{figs/NorthWestFlow.png}
            \caption{Landing geometry of northeast arrivals.}
            \label{fig: NorthEastFlow}
        \end{subfigure}
        \\
        \begin{subfigure}[t]{0.45\textwidth}
            \includegraphics[width=\textwidth]{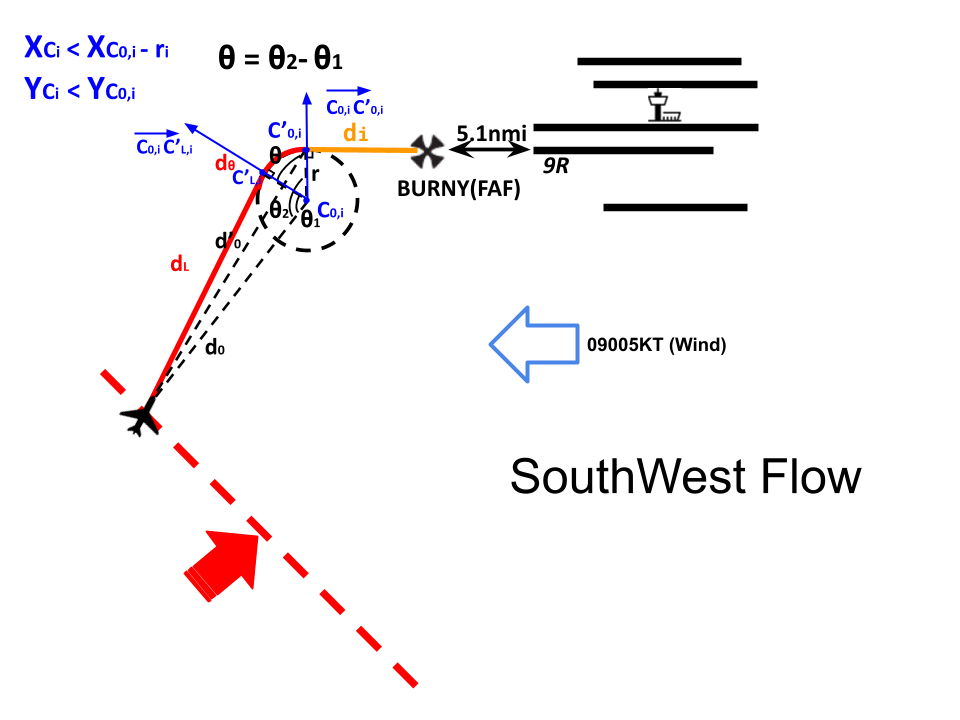}
            \caption{Landing geometry of southwest arrivals.}
            \label{fig: SouthWestFlow}
        \end{subfigure}
        \begin{subfigure}[t]{0.45\textwidth}
            \includegraphics[width=\textwidth]{figs/SouthWestFlow.png}
            \caption{Landing geometry of southeast arrivals.}
            \label{fig: SouthEastFlow}
        \end{subfigure}
    \caption{Arrival path geometry for north and south flows. Each path consists of a tangent leg from the TCP entry fix to the RF turn circle, an RF arc, and a straight-in final segment of length $d_i$ to the FAF. The extension $d_i$ shifts the turn center upstream, increasing total path length and travel time for temporal spacing control. The RF-leg central angle $\theta(d_i)$ is computed from the angle between the two radius vectors at the turn center: one pointing to the tangent point and one pointing to the start of the final straight-in segment.}
    \label{fig: geometry}
\end{figure}

\subsection{TCP Travel Time}
\label{subsec: travel_time}
Let $C_i = (X_{C_i}, Y_{C_i})$ denote the location of aircraft $i$ at the TCP entry time $\tau_i$, and let $(X_{\mathrm{FAF}}, Y_{\mathrm{FAF}})$ denote the coordinates of the Final Approach Fix (FAF). As shown in \Cref{fig: geometry}, the flight path from TCP entry to FAF consists of three segments: (i) a tangent leg of length $d_{L,i}$, flown at speed $v_{L,i}$; (ii) a radius-to-fix (RF) turn arc of length $d_{\theta,i}$ at speed $v_{\theta,i}$; and (iii) a final straight-in segment of length $d_i$ at speed $v_{f,i}$. The extension distance $d_i \ge 0$ is a decision variable representing the trombone (base-leg extension) maneuver upstream of the FAF.

The center of the RF turn circle, $C_{0,i}$, is displaced vertically from the extended FAF position by the turn radius $r$:
\begin{equation}
C_{0,i}(d_i)=
\begin{cases}
\bigl(X_{\mathrm{FAF}}-d_i,\; Y_{\mathrm{FAF}}+r\bigr), & \text{if } Y_{C_i}>Y_{\mathrm{FAF}}+r,\\
\bigl(X_{\mathrm{FAF}}-d_i,\; Y_{\mathrm{FAF}}-r\bigr), & \text{if } Y_{C_i}<Y_{\mathrm{FAF}}-r,
\end{cases}
\end{equation}
where the two cases correspond to arrivals from north and south of the runway centerline, respectively.

\subsubsection{Tangent Leg Length $d_{L,i}$}
Let $d_0(d_i) = \|C_i - C_{0,i}(d_i)\|_2$ denote the Euclidean distance from the entry fix to the turn center. A valid tangent from $C_i$ to the RF circle exists when $d_0(d_i) > r$. By the Pythagorean theorem applied to the right triangle formed by $C_i$, the tangent point, and $C_{0,i}$, the tangent leg length is
\begin{equation}
\label{eq:tangent_leg}
d_{L,i}(d_i) = \sqrt{d_0(d_i)^2 - r^2}.
\end{equation}

\subsubsection{RF Arc Length $d_{\theta,i}$}
The RF arc subtends a central angle $\theta_i(d_i)$ measured from the tangent point $C_{L,i}(d_i)$ on the RF circle to the point $C'_{0,i}(d_i) = (X_{\mathrm{FAF}} - d_i,\, Y_{\mathrm{FAF}})$ where the final straight-in segment begins. The tangent point $C_{L,i}(d_i)$ is the unique point on the circle of radius $r$ centered at $C_{0,i}$ such that $\overrightarrow{C_i C_{L,i}} \perp \overrightarrow{C_{L,i} C_{0,i}}$. Defining the center-to-entry vector $\mathbf{v}_i(d_i) = C_i - C_{0,i}(d_i)$ and its perpendicular $\mathbf{v}_i^\perp(d_i)$, and the coefficients
\begin{equation}
\label{eq:tangent_coeffs}
a_i(d_i) = \frac{r^2}{d_0(d_i)^2}, \qquad b_i(d_i) = \frac{r\sqrt{d_0(d_i)^2 - r^2}}{d_0(d_i)^2},
\end{equation}
the tangent point is given in closed form by
\begin{equation}
\label{eq:tangent_point}
C_{L,i}(d_i) =
\begin{cases}
C_{0,i} + a_i\,\mathbf{v}_i + b_i\,\mathbf{v}_i^\perp, & \text{if } Y_{C_i} > Y_{\mathrm{FAF}} + r, \\
C_{0,i} + a_i\,\mathbf{v}_i - b_i\,\mathbf{v}_i^\perp, & \text{if } Y_{C_i} < Y_{\mathrm{FAF}} - r,
\end{cases}
\end{equation}
where the two cases select the left tangent for north and south arrivals, respectively.

With the left tangent point available, the RF-leg central angle $\theta_i(d_i)$ can be computed directly from the angle between the two radius vectors as
Let $\mathbf{u}_i = \overrightarrow{C_{0,i}C'_{0,i}}$ and $\mathbf{w}_i = \overrightarrow{C_{0,i}C_{L,i}}$ denote the two radius vectors. The RF-leg central angle is
\begin{equation}
\label{eq:theta_arc}
\theta_i(d_i) = \operatorname{atan2}\!\bigl(
\lvert\mathbf{u}_i \times \mathbf{w}_i\rvert,\;
\mathbf{u}_i \cdot \mathbf{w}_i
\bigr)\in[0,\pi].
\end{equation}

Since $\mathbf{u}_i = \overrightarrow{C_{0,i}C'_{0,i}}$ is purely vertical, \eqref{eq:theta_arc} can be written directly in terms of the tangent-point coordinates. Let $\Delta X^L_i = X_{C_{L,i}}(d_i) - X_{C_{0,i}}(d_i)$ and $\Delta Y^L_i = Y_{C_{L,i}}(d_i) - Y_{C_{0,i}}(d_i)$. Then
\begin{equation}
\label{eq:theta_via_CL}
\theta_i(d_i)=
\begin{cases}
\operatorname{atan2}\!\bigl(\lvert\Delta X^L_i\rvert,\; -\Delta Y^L_i\bigr),
& Y_{C_i} > Y_{\mathrm{FAF}}+r,\\[4pt]
\operatorname{atan2}\!\bigl(\lvert\Delta X^L_i\rvert,\;\;\;\Delta Y^L_i\bigr),
& Y_{C_i} < Y_{\mathrm{FAF}}-r,
\end{cases}
\end{equation}
which yields $\theta_i(d_i)\in[0,\pi]$ by construction. The corresponding RF arc length is $d_{\theta,i}(d_i) = r\,\theta_i(d_i)$.

Substituting the closed-form tangent point from \eqref{eq:tangent_point} gives $\theta_i(d_i)$ explicitly in terms of $(X_{C_i}, Y_{C_i})$, $(X_{\mathrm{FAF}}, Y_{\mathrm{FAF}})$, $r$, and $d_i$. Recall
\begin{gather*}
X_{C_{0,i}}(d_i)=X_{\mathrm{FAF}}-d_i,\\
Y_{C_{0,i}}(d_i)=
\begin{cases}
Y_{\mathrm{FAF}}+r, & Y_{C_i} > Y_{\mathrm{FAF}}+r,\\
Y_{\mathrm{FAF}}-r, & Y_{C_i} < Y_{\mathrm{FAF}}-r,
\end{cases}
\end{gather*}
and define
\begin{gather*}
d_0(d_i)=\bigl\lVert C_i - C_{0,i}(d_i)\bigr\rVert_2,
\quad d_0(d_i)>r,\\
a_i(d_i)=\frac{r^2}{d_0(d_i)^2},\quad
b_i(d_i)=\frac{r\sqrt{d_0(d_i)^2-r^2}}{d_0(d_i)^2}.
\end{gather*}
The fully expanded RF-leg central angle is given in~\eqref{eq:theta_expanded}.

To shorten the expanded form, let
$\delta x_i \!=\! X_{C_i}\!-\!X_{\mathrm{FAF}}\!+\!d_i$,
$\delta y_i^{+} \!=\! Y_{C_i}\!-\!Y_{\mathrm{FAF}}\!-\!r$, and
$\delta y_i^{-} \!=\! Y_{C_i}\!-\!Y_{\mathrm{FAF}}\!+\!r$.
Then
\begin{equation}
\label{eq:theta_expanded}
\theta_i(d_i)=
\begin{cases}
\displaystyle\operatorname{atan2}\!\Bigl(
\bigl|\,a_i\,\delta x_i - b_i\,\delta y_i^{+}\bigr|,\;
-\bigl[a_i\,\delta y_i^{+} + b_i\,\delta x_i\bigr]
\Bigr),
& Y_{C_i} > Y_{\mathrm{FAF}}+r,\\[10pt]
\displaystyle\operatorname{atan2}\!\Bigl(
\bigl|\,a_i\,\delta x_i + b_i\,\delta y_i^{-}\bigr|,\;
\phantom{-}\bigl[a_i\,\delta y_i^{-} - b_i\,\delta x_i\bigr]
\Bigr),
& Y_{C_i} < Y_{\mathrm{FAF}}-r,
\end{cases}
\end{equation}
where $a_i = a_i(d_i)$ and $b_i = b_i(d_i)$ for brevity.

\subsubsection{Total Path Length and Travel Time}
The total flight path length is $D_i(d_i) = d_{L,i}(d_i) + d_{\theta,i}(d_i) + d_i$, and the nominal FAF arrival time for aircraft $i$ (ignoring wind) is
\begin{equation}
\label{eq: travel_time}
   t_i = \tau_i + \frac{d_{L,i}(d_i)}{v_{L,i}} + \frac{r\,\theta_i(d_i)}{v_{\theta,i}} + \frac{d_i}{v_{f,i}}
\end{equation}
where $\tau_i$ is the TCP entry time. All geometric quantities ($d_{L,i}$, $\theta_i$, $d_{\theta,i}$) are nonlinear functions of the decision variable $d_i$. The key parameters are the RF turn radius $r$ (we use $r = 2.5$~nmi) and the FAF coordinates (BURNY for runway 9R at KATL). Equation~\eqref{eq: travel_time} is later generalized in \Cref{subsec: wind} to substitute ground speeds for airspeeds under a wind sample.

\subsection{Wind Uncertainty Model}
\label{subsec: wind}
ATC commands the aircraft airspeed (IAS/CAS), but the time at which the aircraft actually reaches the FAF is driven by its ground speed, which differs from the airspeed by the wind component projected onto the aircraft heading \citep{ren2007modeling}. We therefore incorporate the ambient wind model. For each Monte Carlo scenario, a scalar wind speed is sampled from a Gaussian distribution,
\begin{equation}
\label{eq:wind_sample}
w \sim \mathcal{N}(\mu_w,\sigma_w^2),
\end{equation}
with $(\mu_w,\sigma_w) = (5,2)$ knots for the KATL case study, consistent with typical east-to-west surface wind reports in METAR. The wind vector in the coordinate frame used for trajectory geometry (with $x$ pointing east and $y$ pointing north) is $\mathbf{w} = (-w,0)$. Using the horizontal-plane projection in \cite[Sec.~3.2.2 and Appendix~A.3]{ren2007modeling}, for a path segment traveled at airspeed $v$ along a heading unit vector $\hat{\mathbf{h}}$, the effective ground speed along the path is
\begin{equation}
\label{eq:vground}
v_{\mathrm{gs}} = v + \mathbf{w}\cdot\hat{\mathbf{h}} = v - w\,\hat{h}_x,
\end{equation}
which is a tailwind ($v_{\mathrm{gs}} > v$) when the heading has a negative $x$-component, and a headwind otherwise. This is a specialization of Ren's Eq.~(3-1) to the horizontal plane and the small-wind limit ($W \ll V_r$). The three path segments for aircraft~$i$ are treated as follows:
\begin{itemize}[leftmargin=1.2em,itemsep=1pt,topsep=1pt]
  \item \emph{Tangent leg.} The heading is the unit vector from $C_i$ to the tangent point $C_{L,i}(d_i)$. Its $x$-component $\hat{h}_{L,i}(d_i)$ is obtained in closed form from $\bigl(C_{L,i}(d_i)-C_i\bigr)/d_{L,i}(d_i)$ using~\eqref{eq:tangent_point}.
  \item \emph{RF turn.} Because the arc curves, its effective heading is approximated by the chord direction from $C_{L,i}(d_i)$ to the arc-end point $C'_{0,i}(d_i)=(X_{\mathrm{FAF}}-d_i,Y_{\mathrm{FAF}})$; let $\hat{h}_{\theta,i}(d_i)$ denote its $x$-component. This chord approximation is accurate to first order because the RF turn typically sweeps less than $\pi$ radians and the wind-to-airspeed ratio is small.
  \item \emph{Final straight-in segment.} The heading is purely along $+x$ (eastward to the runway), so $\hat{h}_{f,i} = 1$ and the ground speed is simply $v_{f,i} - w$ (pure headwind for the baseline KATL case).
\end{itemize}
Let $w_{L,i}(d_i) = -w\,\hat{h}_{L,i}(d_i)$, $w_{\theta,i}(d_i)=-w\,\hat{h}_{\theta,i}(d_i)$, and $w_{f,i}=-w$ denote the signed wind components along each segment (positive is tailwind). The wind-corrected arrival time is
\begin{equation}
\label{eq:travel_time_wind}
t_i = \tau_i
+ \frac{d_{L,i}(d_i)}{v_{L,i}+w_{L,i}(d_i)}
+ \frac{r\,\theta_i(d_i)}{v_{\theta,i}+w_{\theta,i}(d_i)}
+ \frac{d_i}{v_{f,i}+w_{f,i}},
\end{equation}
which replaces \eqref{eq: travel_time} in the NLP physics constraint and in the MILP arrival-window computation. For NLP tractability, in the constraint Jacobian we evaluate $w_{L,i}(d_i)$ and $w_{\theta,i}(d_i)$ at $d_i = 0$ and treat them as scenario constants; the tangential wind component $w_{f,i}$ is exact and independent of $d_i$. Since discretization and greedy post-processing use the exact ground speed of \eqref{eq:travel_time_wind} for each candidate $d_i$, the NLP approximation has negligible effect on committed solutions.

\subsection{Landing Sequence Determination}
\label{subsec:landing_order}
A critical design choice in arrival scheduling is how the landing order $\Pi$ is determined. The sequence $\Pi$ is a permutation of $\mathcal{I}$ such that $\Pi(k)$ denotes the aircraft index assigned to the $k$-th landing slot. Once the sequence is fixed, it is treated as a hard constraint in the trajectory optimization. We consider two ordering policies:

\subsubsection{First-on-Final-First-Serve (FOFFS)}
Under FOFFS, the landing order is determined by each aircraft's \emph{nominal estimated time of arrival} (ETA) at the FAF, computed assuming zero path extension ($d_i = 0$) and maximum segment speeds:
\begin{equation}
\label{eq:nominal_eta}
\mathrm{ETA}_i^{\mathrm{nom}}
= \tau_i
+ \frac{d_{L,i}(0)}{V_L^{\max}}
+ \frac{r\,\theta_i(0)}{V_\theta^{\max}}.
\end{equation}
Aircraft are sorted by non-decreasing $\mathrm{ETA}_i^{\mathrm{nom}}$, so the aircraft that would arrive earliest at the FAF under free-flight conditions lands first. FOFFS reflects the operational FCFS principle commonly used in terminal airspace: priority is determined by proximity to the runway, not by the order of entry into the TRACON. Because aircraft from different entry fixes travel different path lengths, two aircraft entering the TRACON at the same time generally have different nominal ETAs, and FOFFS naturally exploits this asymmetry to interleave streams and improve throughput.

\subsubsection{First-Entry-First-Serve (FEFS)}
Under FEFS, the landing order is determined strictly by the TCP entry time $\tau_i$. Aircraft are sorted by non-decreasing $\tau_i$, and whoever enters the TRACON first is assigned the earliest landing slot, regardless of path length or nominal ETA. FEFS provides a fairness-oriented baseline that does not exploit geometric asymmetries between arrival streams. Since entry order is observable at the TRACON boundary and does not require any trajectory computation, FEFS represents a simpler operational policy and provides a useful lower bound on the benefit of geometry-aware sequencing.

\subsubsection{Per-Stream Precedence Constraint}
Both FEFS and FOFFS enforce an additional \emph{stream precedence} constraint that is operationally essential in the TRACON environment. Aircraft arriving from the same feeder gate share the same jet route into the terminal area, and traffic controllers do not permit overtaking along a shared jet route \citep{neuman1991analysis, balakrishnan2006scheduling}. Formally, for any two aircraft $i,j\in\mathcal{I}$ that enter from the same corner fix $\kappa$ with $\tau_i < \tau_j$, we require that $i$ must land before $j$ in the final sequence as,
\begin{equation}
\label{eq:stream_precedence}
t_i \le t_j, \qquad \forall (i,j): \kappa_i = \kappa_j,\; \tau_i < \tau_j.
\end{equation}
This precedence constraint has two important consequences. First, it reduces the size of the feasible permutation space. Within the $n!$ possible landing orderings of $n$ aircraft, only those consistent with same-stream monotonicity are feasible. Second, it interacts with CPS: when the window contains multiple aircraft from the same feeder gate, the $y_{ij}$ binaries corresponding to same-stream pairs are fixed in advance and the CPS MILP only reorders across different streams. In practice, this pre-fixing substantially shrinks the effective branch-and-bound tree for CPS$_k$ because a large fraction of the $|\mathcal{S}_k|$ binaries are eliminated.

\subsubsection{Unified Trajectory-Based Scheduling Objective}
\label{subsubsec: unified_objective}
Before introducing the MILP ordering subproblem, we first state the unified objective that is shared by all three sequencing policies evaluated in this paper (FEFS, FOFFS, FOFFS-CPS$_k$) and by both phases of the CPS algorithm. Let $\Pi$ be a landing sequence and let $\{d_i, v_{L,i}, v_{\theta,i}, v_{f,i}, t_i\}_{i\in\mathcal I}$ be the per-aircraft trajectory decision variables; let $\{\sigma_k\}_{k=2}^{N}$ be the separation slacks for consecutive landing ranks. The full trajectory-based arrival scheduling problem is given in~\eqref{eq:unified_obj}, subject to the trajectory constraints (travel-time linking~\eqref{eq:travel_time_wind}, pair-specific separation~\eqref{eq:nlp_sep}, speed monotonicity~\eqref{eq:nlp_speedmono}, and stream precedence~\eqref{eq:nlp_precedence}) and the variable bounds~\eqref{eq:vars_nlp}. The cost penalty weight hierarchy $W_{\mathrm{safe}} \gg W_{\mathrm{thru}} \gg W_{\mathrm{delay}} \ge W_{\mathrm{eff}} \ge W_{\mathrm{speed}}$ encodes the operational priorities: safety first, then throughput, delay, path-stretch efficiency, and a mild speed-maximization regularizer. Here $t_i^{\mathrm{free}} = \tau_i + d_{L,i}(0)/V_L^{\max} + r\,\theta_i(0)/V_\theta^{\max}$ is the free-flight arrival time at $d_i=0$ and maximum speeds. Since $\sum t_i^{\mathrm{free}}$ is constant, the delay term reduces to $W_{\mathrm{delay}}\sum t_i$ in the optimization.

\begin{equation}
\label{eq:unified_obj}
\begin{split}
J^\star \;=\; \min\;\;
& W_{\mathrm{safe}}\!\sum_{k=2}^{N}\!\sigma_k
\;+\; W_{\mathrm{thru}}\,t_N
\;+\; W_{\mathrm{delay}}\!\sum_{i\in\mathcal I}\!(t_i-t_i^{\mathrm{free}})\\
&\;+\; W_{\mathrm{eff}}\!\sum_{i\in\mathcal I}\!d_i
\;+\; W_{\mathrm{speed}}\!\sum_{i\in\mathcal I}\sum_{\ell\in\{L,\theta,f\}}\!\!\frac{V_\ell^{\max}-v_{\ell,i}}{V_\ell^{\max}-V_\ell^{\min}}.
\end{split}
\end{equation}

The unified problem~\eqref{eq:unified_obj} is a mixed-integer nonlinear program (MINLP) with three sources of difficulty: (i)~the integer landing order $\Pi$ over $\mathcal I$, (ii)~the nonlinear trajectory geometry in $d_i$ (via $d_{L,i}(d_i)$, $\theta_i(d_i)$, and the wind-corrected arrival time), and (iii)~the continuous speed variables. Solving the full MINLP for each entering aircraft within the real-time budget of a TRACON scheduler is intractable. We therefore decompose it into a two-phase scheme in which Phase~1 reasons only about the landing order under a scheduling surrogate of $J^\star$, and Phase~2 solves the full nonlinear trajectory optimization given the order fixed by Phase~1. Both phases share the same objective weights and the same safety-first hierarchy but differ only in which decision variables are active.

\subsubsection{FOFFS with Constrained Position Shifting (CPS)}
\label{subsubsec: cps}
FOFFS is a heuristic where the nominal-ETA order is not necessarily optimal when the fleet mix produces large pair-specific gaps (e.g., a Small aircraft behind a Heavy, which incurs the largest entry of \Cref{tab:wake}). We therefore introduce the constrained position shifting policy, denoted FOFFS-CPS$_k$ with the maximum position shift $k$, following the classical CPS formulation in \cite{dear1976dynamic, balakrishnan2006scheduling} but embedded in our online, trajectory-based setting. In the CPS framework, each aircraft is allowed to deviate from its FOFFS rank by at most $k$ positions, where $k\in\{1,2,3\}$ is the range of values recommended in the CPS literature \citep{balakrishnan2006scheduling}, because larger shifts introduce noticeable fairness and controller-workload penalties without commensurate schedule improvements.

Let $\mathcal{W}$ be the set of aircraft in the current online lookahead window, sorted by nominal ETA to obtain a FOFFS rank map $r^{\mathrm{FOFFS}}, \quad \mathcal{W}=\{1,\ldots,|\mathcal{W}|\}$. Let $r^\star$ be the \emph{optimized landing rank} after the MILP is solved. The CPS constraint is,
\begin{equation}
\label{eq:cps_rank}
\bigl|\,r^\star(j) - r^{\mathrm{FOFFS}}(j)\,\bigr| \;\le\; k, \qquad \forall j\in\mathcal{W}.
\end{equation}
Equivalently, an aircraft at FOFFS position $r$ can only land in positions $\{\max(1,r-k),\ldots,\min(|\mathcal W|,r+k)\}$ after resequencing. Implementing \eqref{eq:cps_rank} directly introduces integer position variables so we exploit the fact that any two aircraft whose FOFFS ranks differ by more than $k$ cannot swap relative order, while any two within $k$ of each other can. We therefore partition the window into a wap set and a far set as,
\begin{equation}
\label{eq:cps_sets}
\begin{aligned}
\mathcal{S}_k &= \{(i,j)\in\mathcal W^2:\; 0<r^{\mathrm{FOFFS}}(j)-r^{\mathrm{FOFFS}}(i)\le k\},\\
\mathcal{F}_k &= \{(i,j)\in\mathcal W^2:\; r^{\mathrm{FOFFS}}(j)-r^{\mathrm{FOFFS}}(i) > k\}.
\end{aligned}
\end{equation}
Since Phase~1 must decide only the ordering of aircraft, its active decision variables are the binary ordering indicators $y_{ij}$ and the continuous arrival times $t_j$; the trajectory variables $d_i,v_{L,i},v_{\theta,i},v_{f,i}$ are not available at this phase. We therefore obtain the Phase-1 scheduling surrogate of the unified objective \eqref{eq:unified_obj} by (a) removing the path-stretch term $W_{\mathrm{eff}}\sum d_i$ and the speed regularizer $W_{\mathrm{speed}}$, and (b) replacing the true delay term $W_{\mathrm{delay}}\sum(t_i - t_i^{\mathrm{free}})$ with an excess-delay penalty $W_{\mathrm{delay}}\sum e_j$, where $e_j$ measures the portion of the scheduled delay that cannot be absorbed by speed control alone and must therefore be converted into path stretch $d_j>0$ during Phase~2. This replacement is valid because, for any feasible ordering, the minimum delay achievable by speed control at $d_j=0$ defines an upper bound $E_j + s_j$ above which the Phase-2 NLP is forced to add path stretch. The excess $e_j = \max(0,\,t_j - (E_j + s_j))$ therefore bounds the unavoidable contribution of $\sum d_i$ to $J^\star$.

For each $(i,j)\in\mathcal{S}_k$ we introduce a binary decision $y_{ij}\in\{0,1\}$, where $y_{ij}=1$ means $i$ lands before $j$ and $y_{ij}=0$ means $j$ lands before $i$. For $(i,j)\in\mathcal{F}_k$ we enforce $i$ before $j$ directly (no binary needed). Writing $T_{\mathrm{sep}}^{ij}=T_{\mathrm{sep}}^{\mathrm{wake}}(c_i,c_j)$ for brevity, the resulting Phase-1 MILP is given in~\eqref{eq:milp_p1}, where the objective is the restriction of~\eqref{eq:unified_obj} onto the Phase-1 variable set as described above. The arrival window $[E_j,L_j]$ is computed from~\eqref{eq:travel_time_wind} using $(d_j{=}0,\text{max speeds})$ and $(d_j{=}D_{\max},\text{min speeds})$, respectively. $s_j$ is the delay that speed control alone can absorb at $d_j=0$, $M$ is a large constant, and same-stream aircraft from the same feeder gate are given the precedence $i\!\prec\!j$, if aircraft $i$ entered the gate before $j$. The separation constraints on $\mathcal{F}_k$ use a dedicated slack $\beta_{ij}$ rather than a binary, because the direction is already fixed.

\begin{equation}
\label{eq:milp_p1}
\begin{aligned}
J^\star_{\mathrm{Phase\text{-}1}} \;=\; \min \;& W_{\mathrm{safe}}\!\!\sum_{(i,j)\in\mathcal{S}_k}\!\!\sigma_{ij}
      + W_{\mathrm{safe}}\!\!\sum_{(i,j)\in\mathcal{F}_k}\!\!\beta_{ij}
      + W_{\mathrm{thru}}\,t_{N}
      + W_{\mathrm{delay}}\!\sum_{j\in\mathcal W}\!e_j\\
\text{s.t.}\quad
& E_j \le t_j \le L_j + \alpha_j,\quad j\in\mathcal W,\\
& t_j \ge t_i + T_{\mathrm{sep}}^{ij} - \sigma_{ij} - M(1-y_{ij}),\quad (i,j)\in\mathcal{S}_k,\\
& t_i \ge t_j + T_{\mathrm{sep}}^{ji} - \sigma_{ij} - M\,y_{ij},\quad (i,j)\in\mathcal{S}_k,\\
& t_j \ge t_i + T_{\mathrm{sep}}^{ij} - \beta_{ij},\quad (i,j)\in\mathcal{F}_k,\\
& y_{ij} = 1,\quad \text{if } i,j \text{ same stream and } i \text{ entered first},\\
& e_j \ge t_j - (E_j + s_j),\quad j\in\mathcal W,\\
& t_{N} \ge t_j,\quad j\in\mathcal W,\\
& t_j = t_j^\star,\quad j\in\mathcal W\cap\mathrm{committed},\\
& y_{ij}\in\{0,1\},\; \\&\sigma_{ij},\alpha_j,\beta_{ij},e_j\ge 0.
\end{aligned}
\end{equation}

This formulation closely follows the classical CPS model \citep{balakrishnan2006scheduling, beasley2000scheduling}. The arrival time windows $[E_j,L_j]$, the FCFS/FOFFS rank as the reference ordering, the pair-specific separation matrix, and the same-route precedence constraints are all consistent with the literature. The main differences are, (i) we solve CPS over a sliding online window rather than in static batch; (ii) the separation minima are computed from our pair-specific wake matrix \Cref{tab:wake} combined with runway occupation, and $[E_j,L_j]$ comes from the trajectory-geometry model of \Cref{subsec: travel_time} rather than from jet-route travel times; and (iii) the MILP output is an ordering, which is then passed to Phase~2 that computes actual path stretches and speeds. For reference, the three values of $k$ we consider correspond to \textit{FOFFS-CPS$_1$}, where only adjacent swaps allowed, $|\mathcal{S}_1| = |\mathcal W|-1$ binary variables. \textit{FOFFS-CPS$_2$}, where each aircraft may swap with its two nearest FOFFS neighbors on either side, $|\mathcal{S}_2|\approx 2(|\mathcal W|-1)$ binaries. And \textit{FOFFS-CPS$_3$}, with three-position swaps, $|\mathcal{S}_3|\approx 3(|\mathcal W|-1)$ binaries.

For a window of $n=|\mathcal{W}|$ aircraft, the swap set $\mathcal{S}_k$ contains
$|\mathcal{S}_k|=\sum_{i=1}^{n-1}\min(k,n-i)=kn-k(k+1)/2=\mathcal O(kn)$
ordered pairs, giving $\mathcal O(kn)$ binary variables $y_{ij}$ and $\mathcal O(n^2)$ linear constraints. In the worst case, branch-and-bound on this MILP is exponential in the number of binaries, i.e.\ $\mathcal O(2^{kn})$. This is the same worst-case dependency that Beasley's MILP formulation of the runway scheduling problem exhibits \citep{beasley2000scheduling}, and it is the reason that NP-hardness of the type-dependent CPS problem has been established \citep{balakrishnan2006scheduling}. However, in our scenario of up to more than 60 combined gate arrivals, we maintain a real-time solver performance. The number of distinct orderings of $n$ aircraft with displacement bounded by $k$ from the FOFFS rank is
\begin{equation}
\label{eq:num_cps_orderings}
N_k(n) \;\le\; \prod_{i=1}^{n}(2k+1) \;=\; (2k+1)^n,
\end{equation}
which for $k=1$ grows slowly, while for $k=2$ and especially $k=3$ the bound is loose and the effective combinatorial space grows rapidly. Balakrishnan and Chandran's DP formulation \citep{balakrishnan2006scheduling} exploits this layered structure to obtain a tight $\mathcal O(n\cdot(2k+1)^{2k+1}\cdot(2k+1)!)$ runtime that is polynomial in $n$ but exponential in $k$. Our MILP does not inherit this strong theoretical guarantee, instead, it relies on the LP relaxation plus GLPK's branch-and-cut to close the gap, which is fast in the average case simulated (see~\Cref{sec: solve_time}).

As a practical consequence, increasing $k$ makes the \emph{average} MILP solve only mildly more expensive (Phase-1 time grows roughly linearly with $k$ for $k\!\le\!2$), but the \emph{tail} of the runtime distribution blows up at $k=3$ on symmetric instances. We therefore consider $k\in\{1,2\}$ the practical operational range for the proposed online scheduler. Setting $k=0$ disables the MILP entirely and the scheduler reduces exactly to the FOFFS baseline, which is also numerically verified from our experiments. By construction, FOFFS-CPS$_k$ is a \emph{two-phase} scheme: Phase~1 selects the order; Phase~2 optimizes speeds and path stretch given that order. The online event-driven loop of \Cref{subsec:online} executes both phases each time a new aircraft enters the TRACON.

\subsection{Fixed-Sequence Trajectory Optimization}
\label{subsec:baseline_nlp}
Given a fixed landing sequence $\Pi$ (determined by either FOFFS or FEFS), we formulate a single-stage nonlinear program (NLP) to jointly optimize path extensions and speed profiles for all aircraft in $\mathcal{I} = \{1, \dots, N\}$.

\subsubsection{Decision Variables}

For each $i\in\mathcal{I},$ we optimize the following decision variables,
\begin{equation}
\label{eq:vars_nlp}
\begin{aligned}
& d_i \in [0,D_{\max}],\\
& v_{L,i}\in[V_{L}^{\min},V_{L}^{\max}],\\
& v_{\theta,i}\in[V_{\theta}^{\min},V_{\theta}^{\max}],\\
& v_{f,i}\in\bigl[V_{\text{ref},i}-5,\;V_{\text{ref},i}+20\bigr]\;\text{kts},\\
& t_i\ge 0.
\end{aligned}
\end{equation}
and for each rank $k=2,\dots,N,$ we introduce a soft separation slack variable $\sigma_k \ge 0$. The final-approach speed bounds are \textit{per-aircraft} and are set relative to the type-specific reference landing speed $V_{\text{ref},i}$ (\Cref{subsubsec: fleet}). The lower bound is $V_{\text{ref},i}-5$ knots (a practical deceleration margin toward the runway threshold) and the upper bound is $V_{\text{ref},i}+20$ knots (a representative operational margin for commanded speed on the final segment). This per-aircraft bound replaces the single global interval reflects the fact that a Heavy B773 ($V_{\text{ref}}=150$ kts) cannot realistically fly the same final-approach speed as a Small B735 ($V_{\text{ref}}=127$ kts).

\subsubsection{Optimization Constraints}

\noindent (i) Wind corrected travel-time linking.
The FAF arrival time $t_i$ is linked to path geometry and segment ground speeds via the wind-corrected travel-time formula~\eqref{eq:travel_time_wind}, enforced as an equality constraint for every $i\in\mathcal{I}$. The wind components $w_{L,i},w_{\theta,i},w_{f,i}$ are computed from the scenario wind sample as described in \Cref{subsec: wind}.

\noindent (ii) Pair-specific fixed-sequence separation.
Let $t_k := t_{\Pi(k)}$ denote the FAF arrival time of the aircraft at rank $k$, and let $T_{\mathrm{sep}}^{\,k} = \max\bigl(T_{\mathrm{sep}}^{\mathrm{wake}}(c_{\Pi(k-1)},c_{\Pi(k)}),\allowbreak T_{\mathrm{rwy}}(\Pi(k))\bigr)$ denote the required separation for the corresponding leader--trailer wake-class pair:
\begin{equation}
\label{eq:nlp_sep}
t_k \ge t_{k-1}+T_{\mathrm{sep}}^{\,k}-\sigma_k,
\qquad \forall k\in\{2,\dots,N\},
\end{equation}
where $\sigma_k \ge 0$ allows violations at the cost of a large penalty $W_{\mathrm{safe}}$ in the objective.

\noindent (iii) Speed monotonicity during approach.
Speeds decrease monotonically from the tangent leg to the final approach to ensure a realistic deceleration profile of the aircraft as,
\begin{equation}
\label{eq:nlp_speedmono}
v_{L,i} \ge v_{\theta,i} \ge v_{f,i},
\qquad \forall i\in\mathcal{I}.
\end{equation}

\noindent (iv) Per-stream ordering precedence.
For every pair of aircraft $(i,j)$ originating from the same feeder gate with $\tau_i < \tau_j$, we enforce the non-overtaking constraint \eqref{eq:stream_precedence}:
\begin{equation}
\label{eq:nlp_precedence}
t_j \ge t_i, \qquad \forall (i,j): \kappa_i = \kappa_j,\; \tau_i < \tau_j.
\end{equation}
This constraint is passive for FEFS (the sequence already preserves entry order) but is active for FOFFS and FOFFS-CPS whenever the geometry- or CPS-driven reordering would otherwise cause two same-stream aircraft to swap.

\noindent(v) Variable bounds. The decision variables are bounded as in~\eqref{eq:vars_nlp}, with the per-aircraft final-approach bounds discussed above. The slack variables satisfy $\sigma_k\ge 0$.

\subsubsection{Objective Function}
\label{sec:single_stage_obj}

Given a landing sequence $\Pi$ (produced either directly by FEFS or FOFFS sorting, or by the Phase-1 MILP of \Cref{subsubsec: cps} for FOFFS-CPS$_k$), the Phase-2 trajectory NLP solves the full unified objective over all trajectory decision variables and the separation slacks, subject to the constraints listed above. Formally,
\begin{equation}
\label{eq:nlp_p2}
\begin{split}
J^\star_{\mathrm{Phase\text{-}2}} \;=\; \min\;\;
& W_{\mathrm{safe}}\!\sum_{k=2}^{N}\!\sigma_k
\;+\; W_{\mathrm{thru}}\,t_N
\;+\; W_{\mathrm{delay}}\!\sum_{i\in\mathcal I}\!(t_i-t_i^{\mathrm{free}})\\
&\;+\; W_{\mathrm{eff}}\!\sum_{i\in\mathcal I}\!d_i
\;+\; W_{\mathrm{speed}}\!\sum_{i\in\mathcal I}\sum_{\ell\in\{L,\theta,f\}}\!\!\frac{V_\ell^{\max}-v_{\ell,i}}{V_\ell^{\max}-V_\ell^{\min}}\\[4pt]
\text{s.t.}\quad
& \eqref{eq:travel_time_wind},\;\eqref{eq:nlp_sep},\;\eqref{eq:nlp_speedmono},\;\eqref{eq:nlp_precedence},\;\eqref{eq:vars_nlp},\quad \Pi\;\text{fixed}.
\end{split}
\end{equation}
This is identically $J^\star$ from~\eqref{eq:unified_obj}, now with all five terms active because path stretch $d_i$ and segment speeds are decision variables at this phase. The first term penalizes separation violations, driving $\sigma_k \to 0$ whenever feasible. The second minimizes the makespan $t_N$, promoting throughput. The third minimizes total delay (equivalent to $W_{\mathrm{delay}}\sum t_i$ since $\sum t_i^{\mathrm{free}}$ is a constant). The fourth minimizes total path stretch $\sum d_i$, reducing fuel burn and flight time. The fifth regularizes speeds toward their upper bounds when other objectives are satisfied. The Phase-2 NLP is solved using IPOPT \citep{biegler2009large} via Pyomo.

In summary, both phases solve the same unified objective $J^\star$ with identical weights $W_{\mathrm{safe}}, W_{\mathrm{thru}}, W_{\mathrm{delay}}, W_{\mathrm{eff}}, W_{\mathrm{speed}}$. They differ only in which decision variables are active and in how the path-stretch and speed terms of $J^\star$ are represented. For policies that do not use CPS, Phase~1 is trivial (ordering), Phase~2 is the full NLP formulation, and the two phases together evaluate the same $J^\star$ under the fixed FEFS/FOFFS order.

\subsection{Online Entry Event-Driven Optimization Formulation}
\label{subsec:online}

The static formulation of Section \ref{subsec:baseline_nlp} optimizes all $N$ aircraft as a batch, which requires full knowledge of the arrival set. In practice, aircraft enter the TRACON sequentially and the full demand is not known in advance. We therefore modify the trajectory-based scheduling problem into a rolling-horizon online formulation. At the moment an aircraft enters the TRACON at time $\tau_i$, we solve a finite-horizon subproblem over a lookahead window containing aircraft $i$ and a set of preview aircraft, commit aircraft $i$'s trajectory irrevocably, and advance the event clock to the next entering aircraft. Mathematically, let $\mathcal{C}(t)\subseteq\mathcal{I}$ denote the set of aircraft already committed by time $t$, with their decisions $\{z_j^\star\}_{j\in\mathcal C(t)} = \{(d_j^\star,v_{L,j}^\star,v_{\theta,j}^\star,v_{f,j}^\star,t_j^\star)\}_{j\in\mathcal C(t)}$ held constant for all subsequent subproblems. Let $H(x_i) = \mathrm{ETA}_i^{\mathrm{nom}} - \tau_i$ denote the free-flight time-to-FAF of aircraft $i$ computed from \eqref{eq:travel_time_wind} with $d_i=0$ and maximum speeds. At event $\tau_i$, the online lookahead window is,
\begin{equation}
\label{eq:window}
\mathcal{W}_i \;=\; \mathcal{C}(\tau_i)\;\cup\;\bigl\{\,j\in\mathcal{I}\setminus\mathcal{C}(\tau_i)\,\mid\,\tau_j \le \tau_i + H(x_i)\,\bigr\},
\end{equation}
which consists of all aircraft already committed that still interact with the current scheduling horizon, together with all future aircraft whose TRACON entry time falls within the lookahead. Committed aircraft $j\in\mathcal{C}(\tau_i)\cap\mathcal{W}_i$ enter the subproblem with fixed trajectory variables $z_j = z_j^\star$, and uncommitted aircraft $j\in\mathcal{W}_i\setminus\mathcal{C}(\tau_i)$ carry free decision variables.

The finite-horizon subproblem at event $\tau_i$ is the restriction of the unified optimization to $\mathcal{W}_i$ with the fixed-commitment constraints,
\begin{equation}
\label{eq:subproblem}
\begin{aligned}
\bigl(\Pi^\star_i,\{z_j^\star\}_{j\in\mathcal{W}_i}\bigr)
\;=\;&\arg\min\; J^\star(\mathcal{W}_i)\\
\text{s.t.}\;
& z_j = z_j^\star,\quad j\in\mathcal{C}(\tau_i)\cap\mathcal{W}_i,\\
& \text{with constraints } \eqref{eq:travel_time_wind}\text{--}\eqref{eq:nlp_precedence},\eqref{eq:vars_nlp},\\
& \text{with CPS constraint} \eqref{eq:cps_rank} \text{ if } k\ge 1,
\end{aligned}
\end{equation}
where $J^\star(\mathcal{W}_i)$ denotes the unified objective~\eqref{eq:unified_obj} restricted to the window $\mathcal{W}_i$. Following the two-phase decomposition of \Cref{subsubsec: cps}, we solve~\eqref{eq:subproblem} by first selecting the landing order $\Pi^\star_i$ via the Phase-1 scheduling surrogate, and then computing the full trajectory $\{z_j^\star\}$ via the Phase-2 NLP under the fixed order $\Pi^\star_i$. The resulting online scheduler is summarized in Algorithm~\ref{alg:online}.

\begin{algorithm}[H]
\caption{Online Rolling-Horizon Trajectory-Based Scheduler}
\label{alg:online}
\resizebox{0.85\columnwidth}{!}{%
\begin{minipage}{\columnwidth}
\begin{algorithmic}[1]
\Require Arrival set $\mathcal{I}$ with entry states $\{(x_i,\tau_i,c_i)\}_{i\in\mathcal{I}}$; scenario wind $w$; policy $\pi\in\{\text{FEFS},\text{FOFFS},\text{FOFFS-CPS}_k\}$; objective weights $\mathbf{W}=(W_{\mathrm{safe}},W_{\mathrm{thru}},W_{\mathrm{delay}},W_{\mathrm{eff}},W_{\mathrm{speed}})$; discretization grid $(\Delta_d,\Delta_s)$
\Ensure Committed trajectory $\{z_i^\star\}_{i\in\mathcal{I}}$
\State $\mathcal{C} \gets \emptyset$; $\mathcal{P} \gets \emptyset$ \Comment{committed and preview caches}
\State Sort $\mathcal{I}$ by $\tau_i$ in non-decreasing order
\For{each aircraft $i\in\mathcal{I}$ (in entry-time order)}
    \State $H(x_i)\gets\bigl(d_{L,i}(0)/(V_L^{\max}{+}w_{L,i}){+}r\,\theta_i(0)/(V_\theta^{\max}{+}w_{\theta,i})\bigr)$
    \State Construct lookahead window $\mathcal{W}_i$ via~\eqref{eq:window}
    \If{$i\in\mathcal{P}$ \textbf{and} the cached preview is consistent with $\mathcal{W}_i$ and $\mathcal{C}$}
        \State $z_i^\star \gets \mathcal{P}(i)$ \Comment{reuse preview, skip solve}
    \Else
        \State \textbf{Phase 1 (Sequence Ordering).}
        \If{$\pi = \text{FEFS}$}
            \State $\Pi^\star_i \gets \text{sort}(\mathcal{W}_i,\text{key}=\tau_j)$
        \ElsIf{$\pi = \text{FOFFS}$}
            \State $\Pi^\star_i \gets \text{sort}(\mathcal{W}_i,\text{key}=\mathrm{ETA}_j^{\mathrm{nom}})$
        \ElsIf{$\pi = \text{FOFFS-CPS}_k$, $k\ge 1$}
            \State Solve MILP~\eqref{eq:milp_p1} over $\mathcal{W}_i$ with max-shift $k$
            \State $\Pi^\star_i \gets$ ordering recovered from $\{y_{ij}^\star\}$
        \EndIf
        \State \textbf{Phase 2 (Trajectory Planning).}
        \State Solve NLP~\eqref{eq:nlp_p2} over $\mathcal{W}_i$ with $\Pi=\Pi^\star_i$, fixing $z_j=z_j^\star$ for $j\in\mathcal{C}\cap\mathcal{W}_i$
        \State Round solution to grid $(\Delta_d,\Delta_s)$ via greedy forward-pass repair
        \State $z_i^\star \gets$ committed trajectory of aircraft $i$
        \For{each uncommitted $j\in\mathcal{W}_i\setminus(\mathcal{C}\cup\{i\})$}
            \State $\mathcal{P}(j)\gets z_j^\star$ \Comment{cache preview for future reuse}
        \EndFor
    \EndIf
    \State $\mathcal{C}\gets\mathcal{C}\cup\{i\}$
\EndFor
\State \Return $\{z_i^\star\}_{i\in\mathcal{I}}$
\end{algorithmic}
\end{minipage}%
}
\end{algorithm}

The lookahead window adapts to traffic density tactically. It contains few aircraft under sparse demand, whereas under congestion it grows to approach the size of the static batch subproblem. Two additional properties of Algorithm~\ref{alg:online} are worth noting. First, the \emph{one-touch} commitment rule (line~24) ensures that each aircraft's trajectory is optimized exactly once: once committed, its decision variables $z_j^\star$ enter all subsequent subproblems as constants, so downstream aircraft cannot retroactively change an already-issued clearance. Second, the \emph{preview cache} $\mathcal{P}$ (e.g., lines~6--7, 21--22 of Algorithm~\ref{alg:online}) exploits the fact that the solution for a preview aircraft $j$ is often identical across consecutive subproblems when the local window has not changed. When applicable, it allows Algorithm~\ref{alg:online} to commit $z_j^\star$ without a redundant solver call. The cache-consistency test on line~6 verifies that (i) the current window $\mathcal{W}_i$ is a subset of the window in which $j$ was originally previewed, and (ii) for CPS-enabled policies, no new aircraft has been committed into $\mathcal{C}$ since the preview was stored, since a newly committed aircraft is a hard constraint that could change the Phase-1 ordering. When $k=0$ (FOFFS-CPS$_0$), the Phase-1 MILP is skipped and the scheduler reduces exactly to the FOFFS baseline. 

The NLP~\eqref{eq:unified_obj} produces continuous values of $d_i$ and $v_{L,i},v_{\theta,i},v_{f,i}$. In operational deployment, however, the commanded outputs must be compatible with standard FMS inputs where $d_i$ is expected to take one of a finite set of FAF-extension auxiliary waypoints, and speed commands are typically issued in 5 or 10 knot increments. To study this, we introduce a discretization post-processing step parameterized by $(\Delta_d,\Delta_s)$. After solving the NLP, a greedy forward-pass repair scans aircraft in landing-rank order and, for each uncommitted aircraft, selects the discrete $(d_i,v_{L,i},v_{\theta,i},v_{f,i})$ tuple (from the Cartesian product of $d$-grid and $v$-grid) that minimizes the same weighted-sum objective while satisfying the pair-specific separation from the previously committed leader. When the primary solver is the NLP, this repair step rounds the continuous solution to the nearest feasible grid point; when the primary solver is the discrete greedy, the entire optimization is performed on the $(d,v)$-grid and no continuous step is needed. In either mode the committed output $(d_i^\star,v_{L,i}^\star,v_{\theta,i}^\star,v_{f,i}^\star)$ lies on the prescribed $(\Delta_d,\Delta_s)$ grid, so the results in \Cref{sec: results} directly reflect operationally implementable trajectories.

\subsection{Fuel Consumption Model}
\label{subsec:fuel_consumption}

This subsection describes how fuel burn is calculated for each optimized arrival trajectory using the BADA Total Energy Model (TEM) \citep{nuic_bada_user_manual_36_2004, nuic2010bada} and the calculation described in \cite{ellerbroek2018fuel}. The method (i) compute thrust from TEM, (ii) compute drag from the BADA drag polar, (iii) compute Thrust-Specific Fuel Consumption (TSFC) and fuel flow, and (iv) integrate fuel flow over time.

For each aircraft type (e.g., A321), BADA provides manufacturing parameters and coefficients (from the OPF/APF files), including the reference mass $m$, wing surface area $S$, drag polar coefficients $(C_{D0,\mathrm{cfg}},k_{\mathrm{cfg}})$ per configuration, and fuel coefficients $(C_{f1},C_{f2},C_{f3},C_{f4})$. We use the BADA reference mass when the actual mass is unavailable.

The three trajectory segments optimized in the lateral formulation are paired with a simplified piecewise vertical profile for fuel evaluation. This vertical decomposition follows the constant-CAS idle-descent and level-flight segment structure used in the FMS VNAV path model in \citep[Appendix~A.4.2 and Fig.~A-8]{ren2007modeling}, which in turn reflects the operational RNAV CDA procedures developed in the KSDF flight test series \citep{clarke1997systems, lowther2008enroute} and validated as a baseline CDA concept in \cite{cao2011evaluation}. We simplify their full-CDA profile (including top-of-descent, constant-Mach and deceleration sub-segments) to the following three-segment level-descent-level structure, which is sufficient for fuel evaluation within the TMA,

\begin{itemize}
    \item \textit{Straight-in descent segment ($d_L$).} The aircraft descends from $H_{p,0}=10{,}000~\mathrm{ft}$ to $H_{p,1}=2{,}000~\mathrm{ft}$ at idle thrust while maintaining a constant commanded calibrated airspeed (CAS). This corresponds to the BADA TEM \textit{speed and throttle controlled} mode (control case~a), which is the standard representation of a constant-CAS idle descent, and to the idle constant CAS descent segment in \cite[Sec.~A.4.3]{ren2007modeling}.
    \item \textit{Leveled base-leg RF turn ($d_\theta$).} The RF turn is modeled as level flight at $H_p=2{,}000~\mathrm{ft}$ ($\mathrm{ROCD}=0$) with a commanded constant speed. This matches the BADA TEM \textit{speed and ROCD controlled} mode (control case~c), where thrust is determined from the TEM to maintain the specified speed and ROCD. In this low-altitude portion, the aircraft is assumed to be in approach/landing configuration, so fuel flow is computed from the nominal fuel flow and lower-bounded by the minimum fuel flow, as prescribed by BADA for the approach/landing phases~\citep{nuic_bada_user_manual_36_2004}.
    \item \textit{Leveled final extension ($d_i$).} The final approach extension is similarly modeled as level flight to the FAF at $H_p=2{,}000~\mathrm{ft}$ with a commanded constant speed. As with the base-leg turn, this segment follows the BADA TEM \textit{speed and ROCD controlled} mode with $\mathrm{ROCD}=0$, and fuel flow is computed using the BADA approach/landing fuel-flow rule. Following~\cite{ellerbroek2018fuel}, we assume the FAF altitude is $2,000~\mathrm{ft}$, which coincides with the last fixed vertical constraint in Ren's KSDF CDA example~\citep[Fig.~2-17]{ren2007modeling}.
\end{itemize}

This vertical decomposition is commonly used as a first-order representation of terminal arrival operations and is consistent with the speed-management and deceleration behavior depicted in RNAV arrival/approach profiles~\citep{thomas2021modeling, cao2011evaluation}. The per-aircraft reference landing speeds $V_{\mathrm{ref}}$ in \Cref{tab:ac_params} that size the final-approach speed window play the same role as the final approach speed parameter at the lowest vertical constraint in Ren's FMS VNAV model~\citep[Fig.~A-8]{ren2007modeling}. More detailed vertical procedures, including explicit CDA profiles, configuration schedules, and joint optimization of vertical and lateral procedures, are beyond the scope of the present study and are reserved for future work.

\subsubsection{BADA Total Energy Model}
For aircraft $i$, BADA TEM relates thrust, drag, and the rates of potential/kinetic energy change as,
\begin{equation}
\label{eq:tem}
\mathrm{THR}_i = \frac{m_i g_0}{V_{\mathrm{TAS},i}}\dot h_i + m_i \dot V_{\mathrm{TAS},i} + D_i,
\end{equation}
where $\mathrm{THR}_i$ is thrust along the velocity vector, $m_i$ is aircraft mass, $g_0$ is gravitational acceleration, $V_{\mathrm{TAS},i}$ is true airspeed, $h_i$ is altitude, and $D_i$ is aerodynamic drag. When the true mass is unavailable, we use the BADA reference mass for the corresponding aircraft type.

\subsubsection{Aerodynamic Drag and Configuration}
Drag is computed from dynamic pressure and a configuration-dependent drag polar,
\begin{align}
\label{eq:drag}
D_i &= \frac{1}{2}\rho(h_i)\,V_{\mathrm{TAS},i}^2\,S_i\,C_{D,i},\\
\label{eq:cdpolar}
C_{D,i} &= C_{D0,\mathrm{cfg}} + k_{\mathrm{cfg}}\,C_{L,i}^2,\\
\label{eq:cl}
C_{L,i} &= \frac{2 m_i g_0}{\rho(h_i)\,V_{\mathrm{TAS},i}^2\,S_i\cos\phi_i},
\end{align}
where $\rho(h)$ is air density, $S_i$ is the wing reference area, and $\phi_i$ is bank angle (e.g., $\phi_i=0$ for straight segments). The coefficients $(C_{D0,\mathrm{cfg}},k_{\mathrm{cfg}})$ are selected based on the assumed aerodynamic configuration (e.g., clean/cruise, approach, landing), consistent with BADA's configuration-specific drag definitions. In our terminal-phase fuel evaluation, we use a simplified configuration schedule: the descending straight-in segment uses a clean (or descent) configuration, and the low-altitude level segments at $2{,}000$ ft use an approach configuration.

\subsubsection{Fuel Flow Model for Jet Aircraft}
Following the implementation in \cite{ellerbroek2018fuel}, TSFC is modeled as a function of true airspeed as,
\begin{equation}
\label{eq:tsfc}
\eta(V_{\mathrm{TAS},i}) = C_{f1}\!\left(1+\frac{V_{\mathrm{TAS},i}}{C_{f2}}\right),
\end{equation}
and the nominal fuel flow is linear in TSFC and thrust,
\begin{equation}
\label{eq:fuel_nom}
f_{\mathrm{nom},i} = \eta(V_{\mathrm{TAS},i})\,\mathrm{THR}_i.
\end{equation}
BADA also specifies a minimum (idle) fuel flow as a function of altitude,
\begin{equation}
\label{eq:fuel_min}
f_{\min,i}(h_i)=C_{f3}\!\left(1-\frac{h_i}{C_{f4}}\right).
\end{equation}
In approach/terminal phases, fuel flow is lower-bounded by $f_{\min}$,
\begin{equation}
\label{eq:fuel_limit}
f_i(t)=\max \!\bigl(f_{\mathrm{nom},i}(t),\,f_{\min,i}(h_i(t))\bigr).
\end{equation}
Total fuel burn is obtained by integrating fuel flow over time,
\begin{equation}
\label{eq:fuel_int}
F_i=\int_{t_{i,0}}^{t_{i,f}} f_i(t)\,dt \approx \sum_{j} f_i(t_j)\,\Delta t_j.
\end{equation}

\section{Simulations and Results \label{sec: results}}

\subsection{Experimental Setup}
We evaluate the online scheduling framework through Monte Carlo simulations on the Atlanta A80 TRACON airspace. Arrival traffic from four feeder gates is generated by the shifted Poisson process of \Cref{subsec: flow_generation}, with per-corner rates drawn uniformly from $[1, 30]$~AC/hr, which produces per-scenario aggregate gate arrival rates spanning from 5 up to 60 AC/hr. Each scenario simulates one hour of operations.

Speed bounds reflect realistic approach profiles: $v_{L,i}\in[180,240]$~kts for the tangent leg, $v_{\theta,i}\in[130,200]$~kts for the RF turn, and $v_{f,i}\in[V_{\text{ref},i}-5,V_{\text{ref},i}+20]$~kts per-aircraft on the final segment. The maximum path extension is $D_{\max}=27.5$~nmi. Pair-specific wake separation follows \Cref{tab:wake}, where the minimum same-class gap is $T_{\mathrm{sep}}=66$~s. The NLP is solved with IPOPT via Pyomo with weights $W_{\mathrm{safe}}=10^4$, $W_{\mathrm{thru}}=1$, $W_{\mathrm{delay}}=0.5$, $W_{\mathrm{eff}}=0.1$, $W_{\mathrm{speed}}=0.01$. The MILP for CPS Phase~1 is solved with GLPK.

To study the effect of decision-variable discretization, we sweep a $2\times 2$ grid of path-stretch intervals $\Delta_d\in\{0.5,1.0\}$~nmi and speed intervals $\Delta_s\in\{5,10\}$~kts. The fully-continuous $(\Delta_d=0,\Delta_s=0)$ case is omitted because discrete grids are more operationally actionable (FMS auxiliary-waypoint selection and integer-kt speed commands) and yield essentially the same solution quality after greedy discretization. We evaluate five scheduling policies: \{FEFS, FOFFS, FOFFS-CPS$_1$, FOFFS-CPS$_2$, FOFFS-CPS$_3$\}, giving 20 total configurations. For each configuration, we run $1{,}000$ independent demand seeds and, for each seed, $10$ independent wind samples from $\mathcal{N}(5,2)$ kts, for a total of $200{,}000$ Monte Carlo runs. Unless otherwise stated, all curves report the mean across seeds within a $\pm 2.5$ AC/hr bin and error bars show the combined $\pm 1\sigma$ wind uncertainty.

\subsection{Effect of Ordering Policy on Path Stretch}
\label{sec: effect_order}

\begin{figure}[!t]
    \centering
    \includegraphics[width=\textwidth]{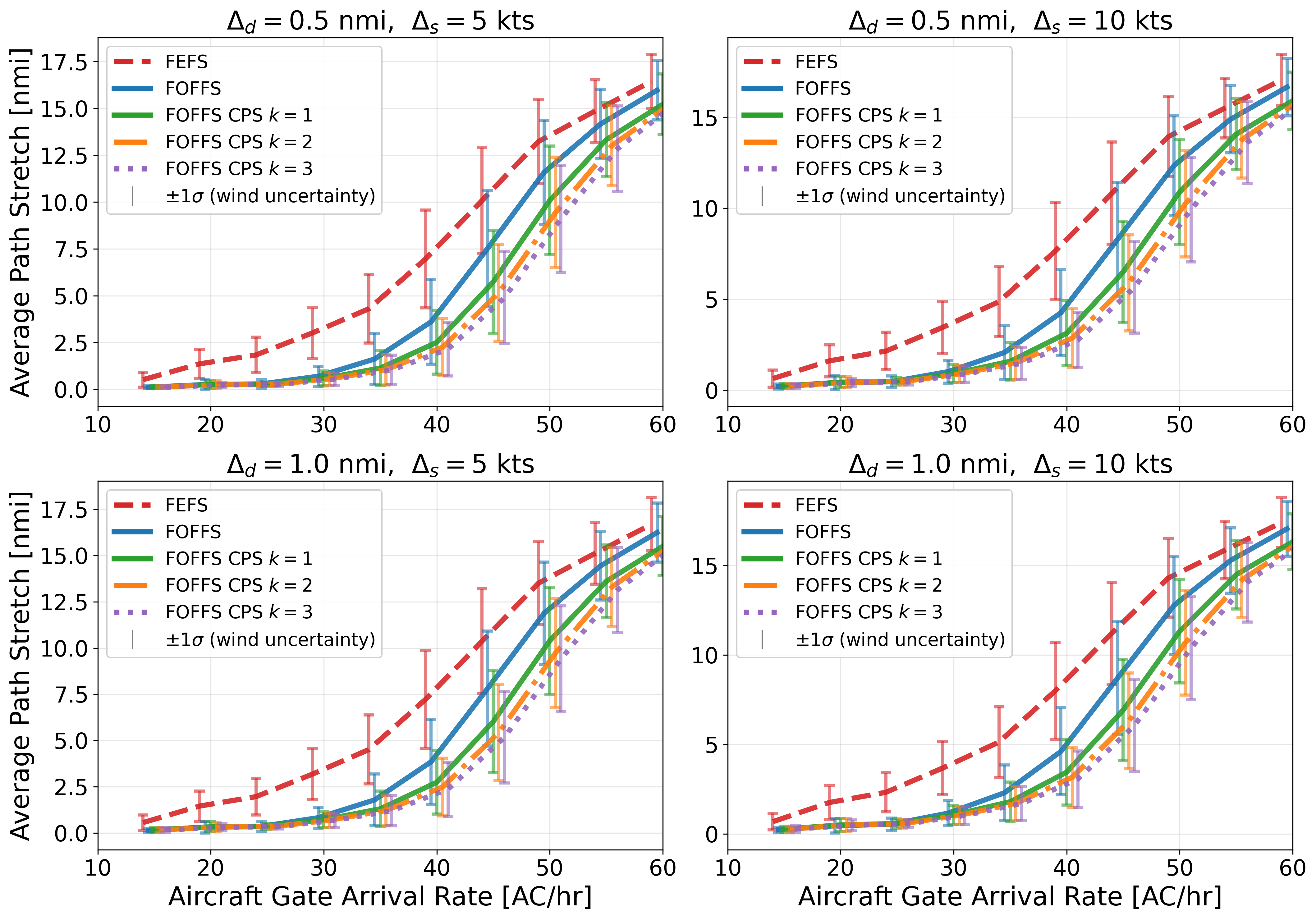}
    \caption{Average path stretch $\bar{d}_i$ vs.\ Aircraft Gate Arrival Rate across the $2\times2$ discretization grid ($\Delta_d$ rows, $\Delta_s$ columns). FEFS (red dashed) always requires the largest extension, while the CPS family ($k\ge 1$) monotonically lowers $\bar{d}_i$ over FOFFS. Error bars show the $\pm 1\sigma$ wind uncertainty from 10 wind samples per seed.}
    \label{fig:di_2x2}
\end{figure}

\Cref{fig:di_2x2} plots the average path stretch $\bar{d}_i$ across the $2\times 2$ discretization grid. Three consistent patterns emerge across all four cells. First, FOFFS achieves substantially lower $\bar{d}_i$ than FEFS at all traffic densities. At low demand FOFFS keeps $\bar{d}_i$ near zero, while FEFS already needs several nmi of extension at around $30$ AC/hr because the entry-time ordering cannot exploit geometric asymmetries between corners. The gap widens with congestion. Second, adding constrained position shifting on top of FOFFS further reduces $\bar{d}_i$. The improvement is monotone but decreasing in $k$ where FOFFS-CPS$_1$ provides the largest single jump over FOFFS, CPS$_2$ adds a smaller benefit, and CPS$_3$ adds only a marginal improvement in most bins. This is consistent with the CPS literature where most of the gain from resequencing comes from swapping adjacent positions to undo the worst leader--trailer pairings (i.e., Heavy-Small), while longer-range swaps rarely pay off once the obvious mispairings have been fixed. Third, the $2\times 2$ grid also isolates the contribution of path-stretch granularity $\Delta_d$ and speed granularity $\Delta_s$. Comparing rows in \Cref{fig:di_2x2}, halving $\Delta_d$ from $1.0$ to $0.5$ nmi reduces $\bar{d}_i$ by $10$--$20\%$ at high demand because the solver can place FAF-extension auxiliary waypoints more finely. Comparing columns, doubling $\Delta_s$ from $5$ to $10$ kts has only a small effect, because the optimizer compensates with the continuous $d_i$ degree of freedom. Overall, path-stretch granularity is the dominant knob where moderate speed quantization (5--10 kts) is operationally attractive without sacrificing solution quality.

The wind-induced variability on $\bar{d}_i$ (the error bars) is much smaller than the inter-seed variability across demand scenarios. For moderate winds ($\mu_w=5$ kts, $\sigma_w=2$ kts) the wind shifts $\bar{d}_i$ by at most $\pm 0.5$ nmi, and the policy ordering remains robust across all wind samples. This validates the per-segment wind projection of \Cref{subsec: wind}.

\subsection{Effect of Ordering Policy on Separation Violation}

\begin{figure}[!t]
    \centering
    \includegraphics[width=\textwidth]{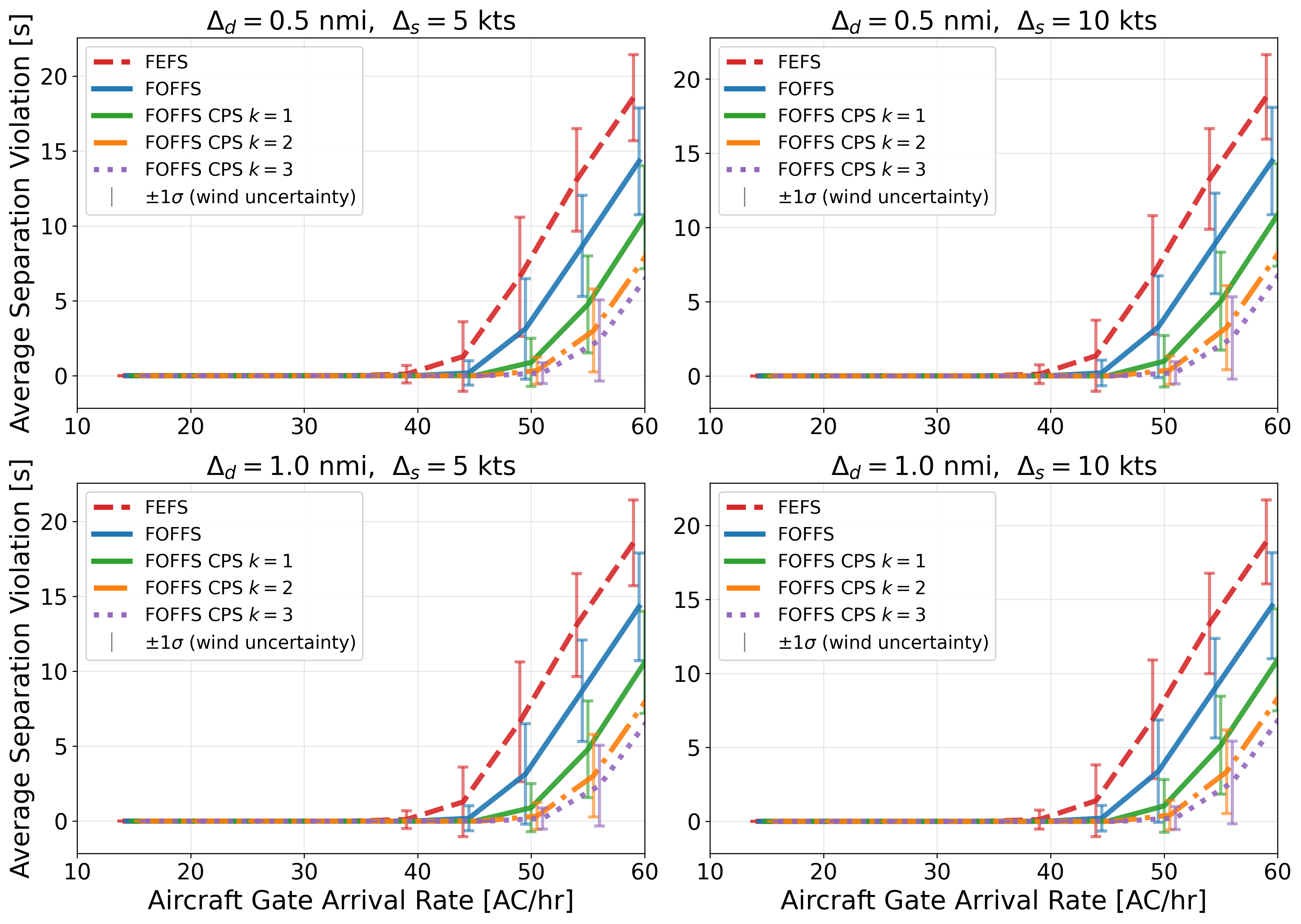}
    \caption{Average separation violation vs.\ Aircraft Gate Arrival Rate. Below $40$ AC/hr all policies maintain essentially zero violation. Above the capacity knee, FEFS climbs the fastest, FOFFS partially compensates through trajectory control, and FOFFS-CPS$_k$ reduces violations further, with diminishing returns from $k=2$ to $k=3$. Note this single runway arrival scenario on a mixed fleet.}
    \label{fig:sep_2x2}
\end{figure}

\Cref{fig:sep_2x2} presents the average separation violation $\bar{\sigma}$ across the same $2\times 2$ grid. The violation metric tells a sharper story than $\bar{d}_i$ because it directly captures where the safety--throughput trade-off starts to fail.

Below $40$ AC/hr, all five policies keep $\bar{\sigma}\approx 0$, confirming that the soft-constraint mechanism with $W_{\mathrm{safe}}=10^4$ drives $\sigma_k\to 0$ whenever physically feasible. Above this, the curves separate dramatically where FEFS climbs the fastest because its entry-order ordering produces many Heavy-Small and other wake-adverse pairings that cannot be undone by path stretch alone. FOFFS partially recovers by letting the solver use trajectory control to soak up the mispairing, but it still runs into infeasibility as the utilization ratio gets closer to 1. The CPS family progressively removes the worst pairings. CPS$_1$ already cuts violations by $50\%$ at 60 AC/hr, CPS$_2$ adds another significant reduction, and CPS$_3$ gives only marginal further improvement.

The capacity near $40$ AC/hr can be explained through a queuing-theoretic lengths. The runway acts as a single server with deterministic service time $T_{\mathrm{sep}}$, while the arrival process is stochastic. The utilization $\rho$ approaches unity well before the deterministic ceiling, and the expected queuing delay grows rapidly to 1. In our setting the queuing delay manifests as increased $\bar{d}_i$ and speed reductions rather than holding-pattern orbits, but the underlying dynamics are the same.

\subsection{Fuel Consumption with BADA vs. OpenAP Cross-Validation}

\begin{figure}[!t]
    \centering
    \includegraphics[width=\textwidth]{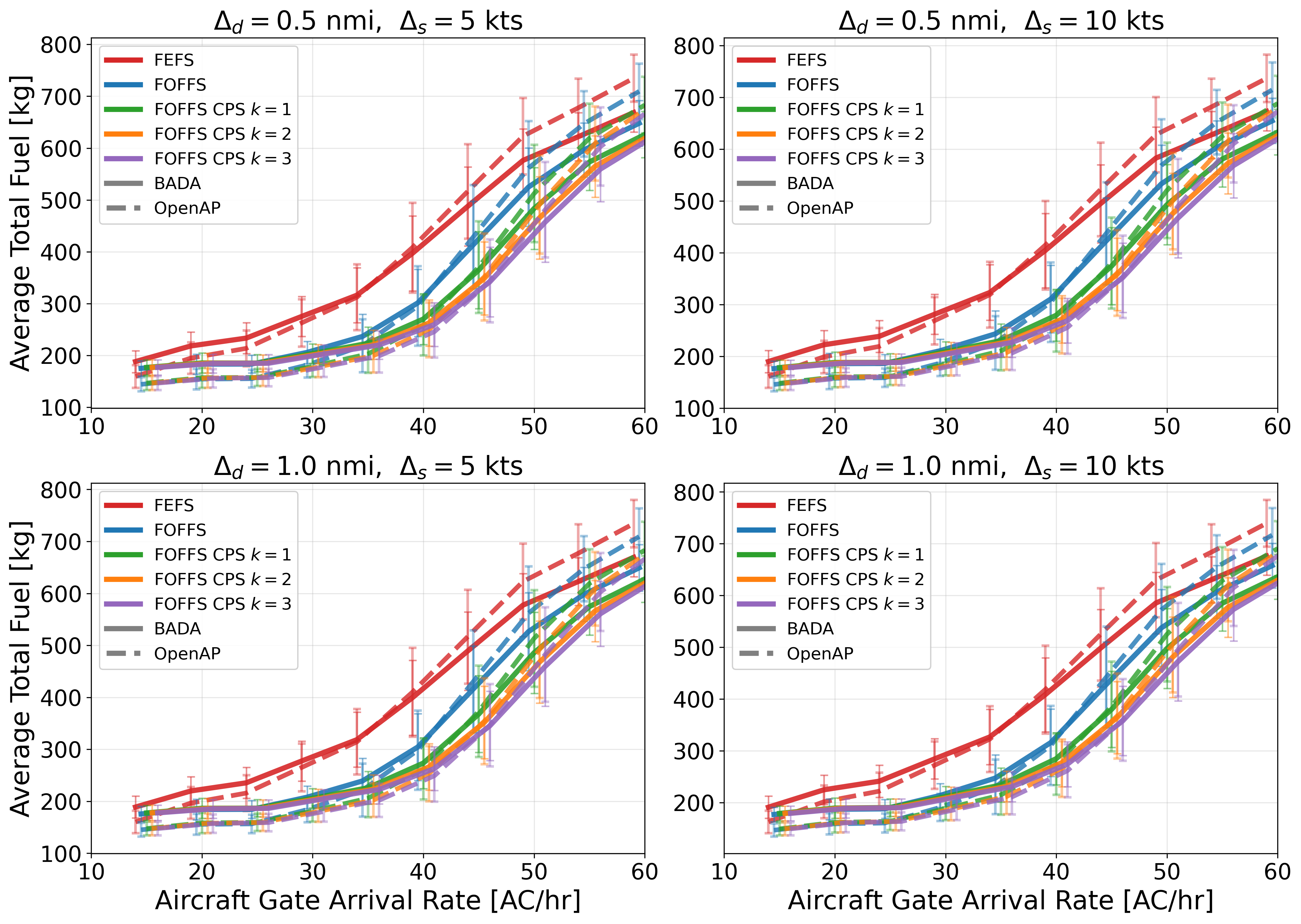}
    \caption{Average per-aircraft fuel burn vs.\ Aircraft Gate Arrival Rate, comparing BADA (solid lines) and OpenAP (dashed lines). Both models agree on the qualitative ordering FEFS $>$ FOFFS $>$ FOFFS-CPS and on the shape of the congestion knee. OpenAP estimates are uniformly slightly higher than BADA due to differences in the drag-polar and fuel-flow parameterization.}
    \label{fig:fuel_2x2}
\end{figure}

\Cref{fig:fuel_2x2} overlays fuel burn estimated by BADA~3~\citep{nuic2010bada} and by the open-source OpenAP performance model~\citep{sun2020openap} under the same discretization grid. Both models reproduce the qualitative ordering FEFS $>$ FOFFS $>$ FOFFS-CPS$_k$ and the same knee location near $45$ AC/hr. OpenAP is uniformly $8\%$ higher than BADA on average (mean $627$ kg vs.\ $581$ kg across all runs), which we attribute to differences in the drag-polar parameterization and the TSFC model. However, the benefit of each policy is identical across the two models. This cross-validation substantially strengthens the fuel-efficiency conclusions, which would otherwise depend on a single parameterization.

Two observations are worth emphasizing. First, at low demand the fuel-burn curves for all five policies are essentially indistinguishable because path stretch is near zero and the baseline segment speeds dominate the fuel integral. Second, at high demand the FEFS curve peels off noticeably above FOFFS, and the CPS curves form a tight bundle slightly below FOFFS where the additional resequencing of CPS$_2$/CPS$_3$ shaves only a small amount of fuel ($10$--$30$ kg per aircraft at 60 AC/hr) because most of the fuel penalty under FEFS comes from the extra path stretch, which is already largely eliminated by plain FOFFS. Wind uncertainty shifts fuel burn by less than $\pm 5\%$, which is below the BADA-OpenAP model gap and below the inter-policy gap at high demand, indicating that the wind uncertainty is a second-order effect for fuel efficiency compared to the ordering policy itself.

\subsection{Real-Time Solver Performance}
\label{sec: solve_time}

\begin{figure}[!t]
    \centering
    \includegraphics[width=\textwidth]{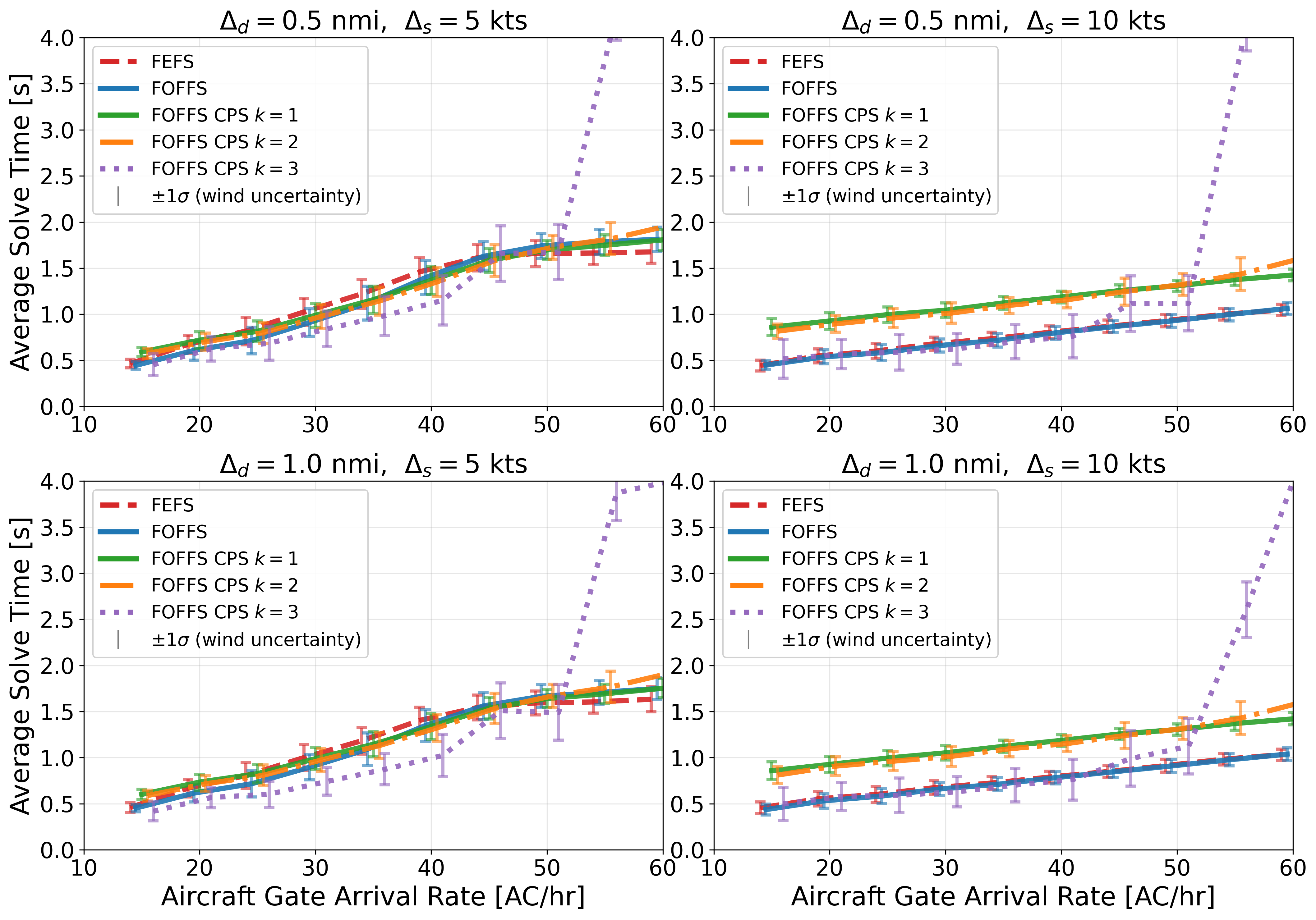}
    \caption{Average per-entry solver runtime vs.\ Aircraft Gate Arrival Rate. FEFS, FOFFS, and FOFFS-CPS$_{1,2}$ remain below $2$~s across the full demand range. Two regimes are visible across the speed-grid columns. At $\Delta_s = 10$~kt (right column) the NLP and discretization-repair pass are cheap, so the constant $\approx 0.4$~s MILP overhead of CPS appears as a parallel offset above the no-MILP FOFFS/FEFS lines. At $\Delta_s = 5$~kt (left column) the finer speed grid makes the per-iteration NLP and repair cost grow super-linearly with window size and dominates the MILP overhead, so all curves merge into a single nonlinear envelope. FOFFS-CPS$_3$ matches the others at low demand but exhibits a heavy-tailed runtime growth above $50$~AC/hr as larger windows admit many feasible permutations and trigger occasional branch-and-bound explosions in the MILP, due to the loose $|\sigma(j)-j|\le 3$ constraint.}
    \label{fig:solve_time_2x2}
\end{figure}

For operational deployment, the online per-entry scheduler must deliver a committed trajectory before the next aircraft enters the TRACON typically within seconds. \Cref{fig:solve_time_2x2} plots the average per-entry wall-clock solver runtime against the aggregate arrival rate.

We have three critical observations. First, FEFS, FOFFS and FOFFS-CPS$_1$/CPS$_2$ all stay below 2 seconds per entry across the full demand range. FOFFS-CPS$_1$ and CPS$_2$ add a small but bounded overhead over FOFFS (a few hundred milliseconds) because the MILP with $k\le 2$ is small and the branch-and-bound tree is shallow. This is comfortably within the real-time budget of an online terminal scheduler. Second, FOFFS-CPS$_3$ matches the others up to $45$ AC/hr, but explodes above $50$ AC/hr where the mean exceeds $6$ s/entry in the highest demand bin, and occasional individual runs take hundreds of seconds. This behavior is consistent with the worst-case complexity analysis in \Cref{subsubsec: cps}. In low-to-medium demand, the lookahead window is small ($n\lesssim 10$) and the number of CPS-compatible orderings is modest. \eqref{eq:num_cps_orderings} gives at most $(2k{+}1)^n$, which for $n=10$, $k=3$ is $7^{10}\approx 2.8\times 10^{8}$, but in practice the LP relaxation and arrival-window cuts close most of the tree in a few iterations. At high demand the window grows to $n\approx 20$--$25$ aircraft, $(2k{+}1)^n$ becomes astronomically large, and the solver becomes sensitive to the tightness of the LP relaxation.

We traced the specific outliers to two concurrent effects. \emph{(i)~Symmetry in the wake matrix.} Several rows of \Cref{tab:wake} contain repeated entries (for example, a leading Small or Large aircraft yields the same $60$~s separation to any trailing Heavy aircraft). When the window contains multiple Heavy aircraft interleaved with Larges, many CPS-valid orderings produce identical makespans within slack tolerance, the LP relaxation is loose, and GLPK's default branching heuristic cannot break the ties. The branch-and-bound tree then enumerates a sizable fraction of the combinatorial many feasible orderings, which pushes the solve time into the seconds-to-minutes range. \emph{(ii)~Wind-induced LP perturbation.} Within the same demand seed, each of the ten wind samples perturbs the arrival-window constants $[E_j, L_j]$ by only a few tenths of a second, but this is sufficient to change which node GLPK selects first in its branching tree. We verified that the resulting schedules are essentially identical (the order and the final committed decisions rarely differ), but the solve times differ by an order of magnitude. This is exactly why the $\pm 1\sigma$ error bars on the CPS$_3$ curve in \Cref{fig:solve_time_2x2} grow dramatically in the high-density bins while the curves for viol, delay, and stretch remain tight.

Two remedies exist for the CPS$_3$ blow-up at high density scenaroios are given here; (a) a commercial MILP solver (CPLEX or Gurobi) with stronger cuts and symmetry breaking typically disposes of these instances in milliseconds, and (b) the dynamic-programming formulation in \cite{balakrishnan2006scheduling} side-steps the symmetry issue entirely by a layered shortest-path search, achieving $\mathcal O(n)$ scaling for each fixed $k$. Either remedy would restore real-time tractability at $k=3$ without affecting the resulting schedule. Within the scope of this paper, we highly regard CPS$_2$ as the practical operating point where the delay/violation reduction from $k=2$ to $k=3$ is modest (\Cref{fig:di_2x2,fig:sep_2x2}), while the solver cost grows by roughly an order of magnitude at high demand. FOFFS-CPS$_1$ is an attractive fall-back when a MILP solver is unavailable, because it captures most of the benefit of resequencing over plain FOFFS at essentially no runtime cost.

\subsection{Discussion}
Taken together, the results demonstrate that the proposed online framework enforces the operational hierarchy embedded in the objective priorities. Safety first, then throughput, then delay and fuel efficiency. Three practical take-aways emerge.

First, FOFFS strictly dominates FEFS because it exploits the natural asymmetry of arrival paths from different entry fixes, and FOFFS-CPS extends this advantage by correcting individual wake-class mispairings at the cost of a small Phase-1 MILP. The gain from resequencing is concentrated in the first few positional shifts. CPS$_1$ already recovers most of the benefit, CPS$_2$ adds a meaningful further reduction, and CPS$_3$ offers only marginal improvement while exhibiting a sharp tail in solver runtime. Operationally, this suggests CPS$_2$ as the recommended working point and CPS$_1$ as an attractive fall-back when a MILP solver is unavailable.

Second, path-stretch granularity $\Delta_d$ is the dominant discretization setup. Halving $\Delta_d$ from $1.0$ to $0.5$ nmi produces a tangible reduction in $\bar{d}_i$ and delay at high demand, whereas doubling the speed granularity $\Delta_s$ from $5$ to $10$ kts has only a small effect because the optimizer compensates through the continuous $d_i$ degree of freedom. This is operationally attractive because discretized $d_i$ maps naturally to a finite set of FAF-extension auxiliary waypoints that can be loaded into the aircraft FMS, while fine speed granularity is not needed.

Third, the online rolling-horizon scheme scales gracefully. At low demand, each windowed NLP/MILP involves only a few aircraft and solves in milliseconds. Under congestion, the window grows but remains bounded by the entering aircraft's free-flight time-to-FAF, and the preview-caching mechanism further reduces redundant computation. The overall wall-clock cost is dominated by the IPOPT-based Phase-2 NLP for FEFS/FOFFS/CPS$_{1,2}$, and only CPS$_3$ at high demand exhibits the adversarial MILP behavior documented in \Cref{sec: solve_time}.

The most important limitation of the present framework is that the vertical profile is decoupled from the lateral optimization and is used only for fuel evaluation (\Cref{subsec:fuel_consumption}). The optimizer has no authority over top-of-descent, deceleration schedule, or glide-slope profile. In reality, these two degrees of freedom are tightly coupled where a later top-of-descent improves cruise efficiency but compresses the usable speed-control envelope on the tangent and turn segments, which in turn shifts the arrival-time window $[E_i,L_i]$ seen by both phases of the scheduler. Vertical--lateral co-optimization is therefore our main planned extension of this work. Building on the CDA and VNAV work of \cite{clarke1997systems, ren2007modeling, lowther2008enroute, cao2011evaluation}, we plan to introduce per-aircraft top-of-descent and deceleration-point decision variables alongside the current lateral $d_i$, and to write the arrival-time constraint as an integral along the full cruise-to-FAF profile as in \cite{gardi2018multiobjective}. Complementary extensions include multiple parallel runways, airline-driven precedence, and a commercial-MILP solver or DP backend to eliminate the CPS$_3$ runtime bottlenecks.

\section{Conclusion \label{sec: conclusion}}
This paper presented a trajectory-based TRACON arrival scheduling framework that couples an analytic terminal path model with an online, entry-driven rolling-horizon optimizer. The main technical contributions of this work are: (i) a closed-form, vector-product derivation of the three-segment arrival path length (tangent leg, radius-to-fix arc, and final straight-in segment) as a smooth nonlinear function of the base-leg extension $d_i$, which exposes the geometric coupling between path stretch and arrival time that node-link abstractions hide and makes the problem directly solvable by an NLP; (ii) a heterogeneous mixed weight class fleet model with pair-specific wake-turbulence separation, type-specific runway occupation, and per-aircraft final-approach speed bounds; (iii) a wind uncertainty model with per-segment heading projection that converts commanded airspeeds into ground speeds; (iv) a FOFFS-with-CPS policy realized by a two-phase online scheme in which a lightweight MILP orders aircraft under a Phase-1 restriction of the unified objective $J^\star$ and a Phase-2 NLP then solves the full $J^\star$ over the trajectory variables under the fixed order; and (v) a large-scale Monte Carlo and real-time solver-runtime analysis spanning nearly $200,000$ runs for rigorous verification of the proposed algorithm.

\section*{Acknowledgment}
{\sloppy This work was supported by the National Aeronautics and Space Administration (NASA) University Leadership Initiative (ULI) program under project \textit{Autonomous Aerial Cargo Operations at Scale}, under grant No.~80NSSC21M071 to the University of Texas at Austin. Any opinions, findings, conclusions, or recommendations expressed in this material are those of the authors and do not necessarily reflect the views of the project sponsor.\par}

\bibliography{ref}

\end{document}